\documentclass [10pt,compsocconf]{IEEEtran}
\usepackage{amsbsy,amsmath,amssymb,epsfig,bbm,mathrsfs,multirow,amsthm}
\usepackage{algpseudocode}

\usepackage{subcaption,float}
\usepackage[hidelinks]{hyperref}

\usepackage[noabbrev]{cleveref}

\crefrangeformat{equation}{#3(#1)#4--#5(#2)#6}

\usepackage[linesnumbered,lined,boxed,ruled,commentsnumbered]{algorithm2e}

\usepackage{blkarray}

\usepackage[T1]{fontenc}
\usepackage[utf8]{inputenc}

\usepackage{caption}

\usepackage[noadjust]{cite}

\newtheorem{lemma}{Lemma}
\newtheorem{theorem}{\textbf{\textsc{Theorem}}}

\newtheorem{corollary}{Corollary}[theorem]

\DeclareMathAlphabet{\mathpzc}{OT1}{pzc}{m}{it}

\DeclareMathOperator*{\argmin}{\arg\!\min}

\begin{document}

\title{Optimal Energy Efficiency with Delay Constraints for Multi-layer Cooperative Fog Computing Networks}


\author{\IEEEauthorblockN{Thai T. Vu, Diep N. Nguyen,  Dinh Thai Hoang, Eryk Dutkiewicz, Thuy V. Nguyen
		\\}\thanks{An abridged version of this paper was presented at the IEEE Globecom Conference, Dec, 2018\cite{vu2018offloading}}}

\maketitle


\begin{abstract}
	We develop a joint offloading and resource allocation framework for a multi-layer cooperative fog computing network, aiming to minimize the total energy consumption of multiple mobile devices subject to their service delay requirements. The resulting optimization involves both binary (offloading decisions) and real variables (resource allocations), making it an NP-hard and computationally intractable problem. To tackle it, we first propose an improved branch-and-bound algorithm (IBBA) that is implemented in a centralized manner. However, due to the large size of the cooperative fog computing network, the computational complexity of the proposed IBBA is relatively high. To speed up the optimal solution searching as well as to enable its distributed implementation, we then leverage the unique structure of the underlying problem and the parallel processing at fog nodes. To that end, we propose a distributed framework, namely feasibility finding
	Benders decomposition (FFBD), that decomposes the original problem into a master problem for the offloading
	decision and subproblems for resource allocation. The master problem (MP) is then equipped with powerful cutting-planes to exploit the fact of resource limitation at fog nodes. The subproblems (SP) for resource allocation can find their closed-form solutions using our fast solution detection method. These (simpler) subproblems can then be solved in parallel at fog nodes. The numerical results show that the FFBD always returns the optimal	solution of the problem with significantly less computation time (e.g., compared with the centralized IBBA approach). The FFBD with the fast solution detection method, namely FFBD-F, can reduce up to $60\%$ and $90\%$ of computation time, respectively, compared with those of the conventional FFBD, namely FFBD-S, and IBBA.
	
\end{abstract}

{\it Keywords-} Task offloading, fog computing, resource allocation, latency, MINLP, branch-and-bound algorithm, and Benders decomposition, distributed computation.


\section{Introduction}
\label{sec:Introduction}

Emerging mobile applications tend to be more and more demanding in computation (e.g., 3D rendering and image processing) as well as lower latency requirements (e.g., interactive games and online object recognition) \cite{soyata2012cloud, rahmani2018exploiting}. Nonetheless, mobile devices are usually limited in computing resources, battery life, and network connections. As a promising solution, a novel network architecture, referred to as mobile edge or fog computing, has recently received paramount interest.
The key idea of fog computing is to ``move'' computing resources closer to mobile users~\cite{Mach2017Mobile}. For that, in a fog computing architecture, powerful computing devices, e.g., servers, are deployed at the edges of the mobile network to support hardware resource-constrained devices, e.g., mobile and IoT devices, to perform high-complexity computational tasks with lower delay. 
Thanks to its unique advantages (e.g., low latency and high bandwidth connections with both mobile devices as well as cloud servers, in the proximity of mobile devices, agile mobility and location awareness support), the fog computing architecture~\cite{hu2015mobile} has proved itself as an effective solution to enable energy-efficient and low-latency mobile applications\cite{mao2017survey}.



However, not all computational tasks benefit from being offloaded to a fog node. Some tasks even consume more energy when being offloaded than being processed locally due to the communication overhead, i.e., sending requests and receiving results \cite{Mach2017Mobile,kumar2010cloud}. 
As such, the task offloading and resource allocation decisions should be jointly considered/optimized. Moreover, unlike  public clouds, e.g., Amazon Web Services and Microsoft Azure, a fog/edge node does not possess abundant computing resource~\cite{el2018edge}. While the computation offloading demand from mobile users is huge, a fog node can support a limited number of tasks. As such the collaboration between fog nodes and cloud servers (referred to as \emph{vertical} collaboration, e.g., ~\cite{chen2018multi,chen2018resource,du2018computation,du2019enabling}) or amongst fog nodes (referred to as \emph{horizontal} collaboration, e.g., ~\cite{xing2019joint,liu2019dynamic,tran2019joint}) is a very promising approach.



Most existing work on joint task offloading and resource allocation of fog computing network only consider either the vertical or horizontal collaboration amongst fog nodes and servers. For example, in \cite{wang2019delay}, a representative of user devices can offload tasks to either nearby fog nodes or a cloud server, aiming to minimize the total delay. However, the authors did not investigate the consumed energy and delay between users and the representative. In~\cite{wang2019cooperative}, the authors developed a joint task offloading and computation
resource allocation scheme to minimize the average task duration. This work did not optimize communication resource allocation since the bandwidth was assumed to be equally shared among associated users. With regard to the horizontal collaboration amongst fog nodes, there are very few works \cite{xing2019joint,liu2019dynamic,tran2019joint}. Authors of \cite{huang2020noma} proposed the joint communication and computation resource sharing scheme in NOMA-aided cooperative computing system. However, a small-scale model with only a user device, a helper, and an access point is considered. Additionally, there is a rich literature on the energy and delay trade-off in fog computing. For that, in this work  we focus on the potential collaboration amongst fog nodes and between fog nodes and the cloud to optimize the decision where is the best for a task to be offloaded to, considering the availability of both computing and communications resources as well as the latency requirement of each task.


Moreover, as aforementioned, although fog nodes' computing capability and resource are more powerful than mobile devices, they are still limited, in comparison with cloud servers. Admitting tasks offloaded from one or a group of mobile devices may prevent it from serving others. Most of the above works and others in the literature tend to overlook this fact, considering only a single-user case. Given the above, this work aims to simultaneously leverage both vertical and horizontal collaboration (amongst fog nodes as well as between fog and cloud server nodes) to jointly optimize the task offloading and resource/computing allocation to minimize the total energy consumption of multiple mobile users (subject to their diverse latency requirements). To that end, we introduce a multi-layer mixed fog and cloud computing system including multiple users, multiple fog nodes, and the remote cloud server. The model allows us to exploit the advantages of both fog nodes and the cloud server as well as enables the scalability due to the collaboration between fog nodes. To overcome the drawbacks of~\cite{wang2019delay}~and~\cite{wang2019cooperative}, we investigate all factors (i.e., uplink, downlink and processing) contributing to the overall delay, consumed energy as well as communication resource allocation.

The resulting optimization involves both binary (offloading decisions) and real variables (resource allocations), called a mixed integer non-linear programming problem (MINLP). That makes it an NP-hard and computationally intractable problem. To tackle it, we first propose an improved branch-and-bound algorithm (IBBA) that is implemented in a centralized manner. However, due to the large size of the cooperative fog computing network, the computational complexity of the proposed IBBA is relatively high. To speed up the optimal solution searching as well as to enable its distributed implementation, we then leverage the unique structure of the underlying problem and the parallel processing at fog nodes. To that end, we then propose a distributed approach, namely feasibility finding
Benders decomposition (FFBD), that decomposes the original problem into a master problem for the offloading
decision and subproblems for resource allocation. The master problem (MP) for the offloading decisions is then equipped with powerful cutting-planes based on resource limitation of fog nodes. The subproblems (SP) for resource allocation can find closed-form solutions using our fast solution detection method. These (simpler) subproblems can be then solved in parallel at fog nodes. The numerical results show that the FFBD always returns the optimal	solution of the problem with significantly less computation time (e.g., in comparing with the centralized approach). The FFBD with the fast solution detection method can reduce up to $60\%$ and $90\%$ of computation time, respectively, than those of the conventional FFBD and IBBA. %
Major contributions of this paper are as follows:

%
%
%

\begin{itemize}
	
	
	\item We propose a cooperative computing framework which considers both vertical and horizontal collaboration amongst fog and cloud nodes while minimizing the total energy of all mobile devices, subject to their service delay constraints.
	
	

	\item We then propose an improved branch-and-bound (IBBA) method that exploits special features of our task offloading model to obtain different optimal offloading policies.

	\item 
	To leverage the computation capability at all fog nodes, we develop a distributed feasibility-finding Benders decomposition (FFBD) algorithm. The algorithm decomposes the original problem into a master problem (MP) for offloading decisions and multiple subproblems (SP) for resources allocation. Exploiting special characteristics of the problem, the subproblems (SP) in the FFBD can be solved independently at edge nodes.
	
	
	\item To further reduce the computation time of FFBD, we develop a theoretical framework for the feasibility and infeasibility detection of the subproblems based on fog nodes' resource limitation. Then, the master problem is equipped with powerful cutting-planes using theoretical analysis on the infeasibility of SPs. The subproblems can then find closed-form solutions using the fast solution detection method.	
	
	
		
	
	\item We perform intensive simulations to evaluate the efficiency of the proposed framework and solutions, and compare their performance (i.e., the consumed energy, delay, processing time and complexity) with those of the standard solutions. These results provide insightful information on factors affecting the performance of the proposed methods.

\end{itemize}

The rest of the paper is as follows. We describe the system model and the problem formulation in Section~\ref{sec:sysmodel}. Section \ref{sec:solutions} presents the proposed algorithms (IBBA, FFBD-S, FFBD-F with different optimal solution selection criteria), the theoretical analyses, the optimal solution selection strategies. In this section, we also design a protocol to implement the proposed algorithms. In Section \ref{sec:performanceevaluation}, we evaluate the performance of proposed algorithms and compare them with different baseline methods. 
Conclusions are drawn in Section \ref{sec:conclusion}.

\section{System Model and Problem Formulation}
\label{sec:sysmodel}

\subsection{Network Model}
\begin{figure}[!]
	\centering
	\includegraphics[scale=0.35]{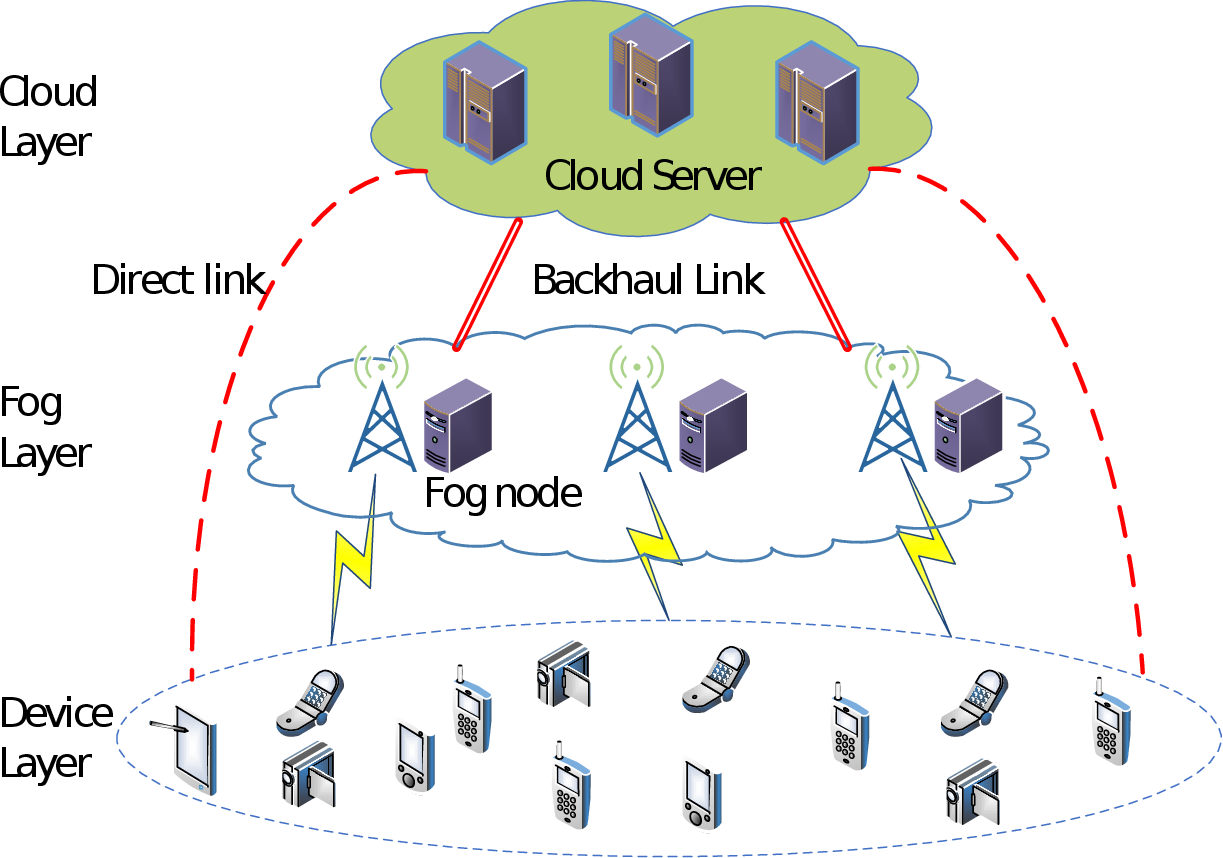}
	\caption{Three-layer cooperative fog computing network.}
	\label{fig:System-Model}
\end{figure}

Fig.~\ref{fig:System-Model} illustrates a multi-layer fog computing system with $N$ mobile devices $\mathbb{N}=\{1,\ldots,N\}$, $M$ cooperative fog nodes $\mathbb{M}=\{1,\ldots,M\}$, and a cloud server $V$ that can be directly reached by all mobile devices (e.g., the cloud server is co-located with the base station). A mobile device can offload a task to either one of fog nodes, the cloud via a fog node or directly to the cloud server $V$. To focus on the impact of multi-user on the offloading and resource allocation decisions, we assume that a multi-access method is in place at the MAC layer, e.g., OFDMA or NOMA. 

At each time slot, mobile user $i$ can request to offload a{\footnote{The following analysis also applies if a mobile user has multiple tasks at the same time}} computing task  $I_{i}\left(D_{i}^{i},D_{i}^{o},C_{i},t_{i}^{r}\right)$, in which $D_{i}^{i}$ and $D_{i}^{o}$ respectively are the input (including input data and execution code) and output/result data lengths, $C_{i}$ is the number of CPU cycles that are required to execute the task, and $t_{i}^{r}$ is the maximum delay requirement of the task. Only the mobile device, fog nodes, or the cloud server satisfying the delay requirement are eligible to process the task.

Under the multi-tier/layer model, the highest tier is the cloud server that has the highest CPU rate but would require more energy and latency for the mobile nodes to offload to. On the other hand, the second tier, i.e., fog nodes, has lower CPU rates than the cloud's but is closer to mobile devices. As such, in terms of communications latency and energy, fog nodes are more preferable than the cloud for mobile devices to offload on. The last tier is the mobile devices who have the lowest CPU rates but if they select to process the tasks locally, then the communications latency is the lowest or zero, compared with offloading the tasks to the fog or the cloud.

\subsubsection{Local Processing}

Mobile device $i$ has a processing rate $f_{i}^{l}$ in cycles per second.
If task $I_i$ is processed locally, the time to perform the task is given by
\begin{equation}
T_{i}^{l}=C_{i}/f_{i}^{l}.
\label{eq:local_delay}
\end{equation}

The consumed energy $E_{i}^{l}$ of the mobile device is proportional to the CPU cycles required for task $I_{i}$ and is given by
\begin{equation}
E_{i}^{l}=v_{i}C_{i},
\label{eq:local_en}
\end{equation}
where $v_{i}$ denotes the consumed energy per CPU cycle~\cite{Wen2012Energy,chen2018task}.


%

\subsubsection{Fog Node Processing}
Fog node $j$ has capabilities denoted by a tuple $(R_{j}^{u},R_{j}^{d},R_{j}^{f})$ in which $R_{j}^{u}$, $R_{j}^{d}$, and $R_{j}^{f}$ are the total uplink rate, total downlink rate, and CPU cycle rate respectively. 
If task $I_i$ is processed at fog node $j$, then the fog node will allocate radio/communications and computation resources for the mobile device/task $I_i$, defined by a tuple $\mathbf{r}_{ij}=(r_{ij}^{u},r_{ij}^{d},r_{ij}^{f})$, in which $r_{ij}^{u}$, $r_{ij}^{d}$, $r_{ij}^{f}$ respectively are uplink, downlink, and CPU cycle rates for input, output transmissions, and executing the task. In this case, the energy consumption at the mobile user is for both transferring input to and receiving output from fog node $j$, and the delay includes time for transmitting input, receiving output and task processing at the fog node.

Various physical factors (i.e., channel fading, antenna gain, power level, circuit power, and active/receive time of transmitter/modem) affect energy efficiency of communication at each mobile device. These parameters are readily available at the physical layer can be captured through the energy required per up-/down-link (transmit/receive) bit. Note that the transmitting/receiving time at fog/cloud is relatively small compared with the channel coherence time so that the channel state can be assumed to be unchanged during transmission. Let $e_{ij}^{u}$ and $e_{ij}^{d}$ denote the energy consumption for transmitting and receiving a unit of data between the mobile device $i$ and the fog node $j$, respectively. The delay $T_{ij}^{f}$ and the consumed energy $E_{ij}^{f}$ of mobile device are given by:
\begin{equation}
T_{ij}^{f}=D_{i}^{i}/r_{ij}^{u}+D_{i}^{o}/r_{ij}^{d}+C_{i}/r_{ij}^{f},
\label{eq:fog_delay}
\end{equation}
and
\begin{equation}
E_{ij}^{f}=E_{ij}^{u}+E_{ij}^{d},
\label{eq:fog_en}
\end{equation}
where $E_{ij}^{u}=e_{ij}^{u}D_{i}^{i}$ and $E_{ij}^{d}=e_{ij}^{d}D_{i}^{o}.$

\subsubsection{Cloud Server Processing (offloaded via a fog node)}
Without loss of generality, we can assume that all fog nodes are connected to a public cloud server. We denote the backhaul data rate between a fog node and the cloud server as $w^{c}$, and the processing rate assigned to each task on the cloud server as $f^{c}$.


If fog node $j$ forwards task $I_i$ to the cloud server, it will allocate resources for mobile device~$i$, defined by a tuple $\mathbf{r}_{ij}=(r_{ij}^{u},r_{ij}^{d},r_{ij}^{f})$, in which $r_{ij}^{u}$, $r_{ij}^{d}$ are uplink rate, downlink rate for input and output transmissions, and $r_{ij}^{f}=0$. After receiving the task, fog node $j$ sends the input data to the cloud server for processing, then receives and sends the result back to the mobile user. 
In this case, the consumed energy $E_{ij}^{c}$ at the mobile user is only for transmitting  input and output data directly to and from fog node $j$ as in the case of fog node processing, while the delay $T_{ij}^{c}$ includes the time for transmitting the input from mobile user to the fog node, time from the fog node to the cloud server, time for receiving the output from the cloud server to mobile user via the fog node, and task-execution time at the cloud server. These performance metrics are as follows:
\begin{equation}
T_{ij}^{c}=D_{i}^{i}/r_{ij}^{u}+D_{i}^{o}/r_{ij}^{d}+
(D_{i}^{i}+D_{i}^{o})/w^{c}+C_{i}/f^{c},
\label{eq:cloud_delay}
\end{equation}
and
\begin{equation}
E_{ij}^{c}=E_{ij}^{f}=E_{ij}^{u}+E_{ij}^{d}.
\label{eq:cloud_en}
\end{equation}

\subsubsection{Cloud Server Processing (directly offloaded by mobile devices)}


Let $\mathbb{M^*} = \mathbb{M} \cup \{V\}$ be the set of $M$ fog nodes and the cloud server $V$. The cloud $V$ can be now considered as the $(M+1)$th node of $\mathbb{M^*}$.  The direct connection from mobile devices to the cloud is captured by the tuple $(R_{(M+1)}^{u},R_{(M+1)}^{d},R_{(M+1)}^{f})$ in which $R_{(M+1)}^{u}$, $R_{(M+1)}^{d}$, and $R_{(M+1)}^{f}$ are uplink capacity, downlink capacity, and the CPU cycle rate of the cloud, respectively. We have $R_{(M+1)}^{f}=f^c$.  

For a task $I_i$ to be directly offloaded and executed by the cloud $V$ (or the $(M+1)$ node), the cloud will allocate radio/communications and computation resources, defined by a tuple $\mathbf{r}_{i(M+1)}=(r_{i(M+1)}^{u},r_{i(M+1)}^{d},r_{i(M+1)}^{f})$, in which $r_{i(M+1)}^{u}$, $r_{i(M+1)}^{d}$, $r_{i(M+1)}^{f}$ respectively are the cloud's uplink, downlink, and CPU cycle rates for input, output transmissions, and executing the task. The corresponding delay $T_{i(M+1)}^{f}$ and energy consumption $E_{i(M+1)}^{f}$ for this case can then be calculated similarly to those in equation (3) and (4):

\begin{equation}
T_{i(M+1)}^{f}=D_{i}^{i}/r_{i(M+1)}^{u}+D_{i}^{o}/r_{i(M+1)}^{d}+C_{i}/r_{i(M+1)}^{f},
\label{eq:direct_cloud_delay}
\end{equation}
and
\begin{equation}
E_{i(M+1)}^{f}=E_{i(M+1)}^{u}+E_{i(M+1)}^{d},
\label{eq:direct_cloud_en}
\end{equation}
where $E_{i(M+1)}^{u}=e_{i(M+1)}^{u}D_{i}^{i}$ and $E_{i(M+1)}^{d}=e_{i(M+1)}^{d}D_{i}^{o}$ (with that $e_{i(M+1)}^{u}$ and $e_{i(M+1)}^{d}$ are the energy consumption for transmitting and receiving a unit of data between the mobile device $i$ and the cloud).

Note that for the cloud (i.e., node $(M+1)$), as itself is the top/last tier, it can't choose to offload a task to a higher tier. This is mathematically captured by setting $T^c_{i(M+1)}=\infty$ and $E_{i(M+1)}^{c}$ as a constant.

\subsection{Problem Formulation}
\label{sec:problemformulation}


We denote the binary offloading decision variable for task $I_{i}$ by $\mathbf{x}_{i}=(x_{i}^{l},x_{i1}^{f}, \dots,x_{i(M+1)}^{f},x_{i1}^{c}, \dots,x_{i(M+1)}^{c})$, in which $x_{i}^{l}=1$, $x_{ij}^{f}=1$, and $x_{ij}^{c}=1$ respectively indicate that task $I_{i}$ is processed at the mobile device, fog node $j$ (or directly at the cloud if $j=M+1$), or the cloud server (via fog node $j$ for $j<M+1$). As mentioned above, the cloud $V$ or the node $(M+1)$th is the top-tier, so it can't offload a task to a higher tier or $x_{i(M+1)}^{c}=0$.  

Let $\mathbf{h}_i = (T_i^l, T_{i1}^f,\dots,T_{i(M+1)}^f,T_{i1}^c,\dots,T_{i(M+1)}^c)$. From Eqs.~(\ref{eq:local_delay}),~(\ref{eq:fog_delay}),~(\ref{eq:cloud_delay}),~and~(\ref{eq:direct_cloud_delay}), the delay $T_i$ when task $I_i$ is processed is given as
\begin{equation}
T_{i}=\mathbf{h}_i^{\top}\mathbf{x}_{i}.
\label{eq:task_delay_nonconvex}
\end{equation}

Note that the delay $T_i$ in Eq.~(\ref{eq:task_delay_nonconvex}) is not a convex function w.r.t. the offloading and resource allocation decisions. This is due to the its non-convex components in the form of the ratio of these variables $x/r$ (where $x$ and $r$ are offloading decision and resource allocation variables, respectively). Consequently, the formulated problem with delay constraints is not a convex optimization problem. To leverage convex optimization, we convert $T_i$ in Eq.~(\ref{eq:task_delay_nonconvex}) to a convex one. Because $x_{i}^{l},x_{ij}^{f},x_{ij}^{c}$ are binary variables, we have $x_{i}^{l}=(x_{i}^{l})^2,x_{ij}^{f}=(x_{ij}^{f})^2,~\text{and}~ x_{ij}^{c}=(x_{ij}^{c})^2$. Thus, we can equivalently reformulate $T_i$ as
\begin{equation}
T_{i}=\mathbf{h}_i^{\top}\mathbf{y}_{i},
\label{eq:task_delay_convex}
\end{equation}
where $\mathbf{y}_{i}=((x_{i}^{l})^2,(x_{i1}^{f})^2, \dots,(x_{i(M+1)}^{f})^2,(x_{i1}^{c})^2, \dots,(x_{i(M+1)}^{c})^2)$.

In the rest of the paper we will use $T_i$ described in Eq.~(\ref{eq:task_delay_convex}) for the delay of task $I_i$. The convexity of $T_i$ in Eq.~(\ref{eq:task_delay_convex}) will be proven and used in Theorem~\ref{theo:convexity}.

Let $\mathbf{e}_i = (E_i^l, E_{i1}^f,\dots,E_{i(M+1)}^f,E_{i1}^c,\dots,E_{i(M+1)}^c)$. From Eqs.~(\ref{eq:local_en}),~(\ref{eq:fog_en}),~(\ref{eq:cloud_en}),~and~(\ref{eq:direct_cloud_en}), the consumed energy $E_i$ of the mobile user when task $I_i$ is processed are given as
\begin{equation}
E_{i}= \mathbf{e}_{i}^{\top}\mathbf{x}_{i}.
\label{eq:task_en}
\end{equation}

Let $\mathbf{e} = (\mathbf{e}_1, \dots, \mathbf{e}_N)$ and $\mathbf{x} = (\mathbf{x}_1, \dots, \mathbf{x}_N)$. The total consumed energy of mobile devices is given as
\begin{equation}
E= \mathbf{e}^{\top}\mathbf{x}.
\label{eq:total_en}
\end{equation}

In this paper, we address a joint offloading decision $(\mathbf{x})$ and resource allocation $(\mathbf{r}=\{\mathbf{r}_{ij} \} )$ problem that aims to minimize the total energy consumption of all mobile devices under the delay requirement. The problem is formally stated as follows.
\begin{equation}
(\mathbf{P}_0) \hphantom{10}	 \underset{\mathbf{x},\mathbf{r}}{\min} ~\mathbf{e}^{\top}\mathbf{x} ,
\label{eq:global_en}
\end{equation}
s.t.
\begin{equation}
(\mathbf{R}_0)
\begin{aligned}
\left\{	\begin{array}{ll}
(\mathcal{C}_1) \hphantom{1} T_{i} \leq t_{i}^{r}, \forall i\in \mathbb{N},	\\
(\mathcal{C}_2)  \hphantom{1} \sum_{i=1}^{N}r_{ij}^{f} \leq R_{j}^{f}, \forall j\in \mathbb{M^*},	\\
(\mathcal{C}_3)  \hphantom{1} \sum_{i=1}^{N}r_{ij}^{u} \leq R_{j}^{u}, \forall j\in \mathbb{M^*},	\\
(\mathcal{C}_4)  \hphantom{1} \sum_{i=1}^{N}r_{ij}^{d} \leq R_{j}^{d}, \forall j\in \mathbb{M^*},	\\
r_{ij}^{u}, r_{ij}^{d}, r_{ij}^{f} \geq 0,\forall (i,j)\in \mathbb{N}\times\mathbb{M^*},
\end{array}	\right.
\end{aligned}
\label{eq:resource_delay_con}
\end{equation}
and
\begin{equation}
(\mathbf{X}_0)
\begin{aligned}
\left\{	\begin{array}{ll}
(\mathcal{C}_5) \hphantom{1} x_{i}^{l}+\sum \limits_{j=1}^{M+1}x_{ij}^{f}+\sum \limits_{j=1}^{M+1}x_{ij}^{c}=1, \forall i\in \mathbb{N},	\\

x_{i}^{l},x_{ij}^{f}, x_{ij}^{c} \in\{0,1\}, \forall (i,j)\in \mathbb{N}\times\mathbb{M^*}.

\end{array}	\right.
\end{aligned}
\label{eq:offload_qos_con}
\end{equation}
where $(\mathcal{C}_1)$ is the delay requirement of tasks, $(\mathcal{C}_2), (\mathcal{C}_3)~\text{and}~ (\mathcal{C}_4)$ are resource constraints at fog nodes, and $(\mathcal{C}_5)$ is offloading decision constraints.

\section{Proposed Optimal Solutions}
\label{sec:solutions}
The optimization problem $(\mathbf{P}_0)$ is an NP-hard due to its mixed integer non-linear programming. We observe that by relaxing all its binary variables to real numbers $x_{i}^{l},x_{ij}^{f}, x_{ij}^{c} \in [0,1], \forall (i,j)\in \mathbb{N}\times\mathbb{M^*}$, the resulting problem (referred to as the fully-relaxed problem) is a convex optimization problem~\cite{Boyd2004Convex}. The convexity of the fully-relaxed problem is maintained in partly-relaxed problems, that are obtained by fixing some binary variables (to be $0$ or $1$) and relaxing the remaining ones. Using this characteristic, in the sequel we introduce three effective approaches to address the problem $(\mathbf{P}_0)$.

\subsection {Convexity of Relaxed Problems}
The fully-relaxed problem is written as follows: 
\begin{equation}
(\mathbf{\widetilde{P}}_0) \hphantom{10}	 \underset{\mathbf{x},\mathbf{r}}{\min} ~\mathbf{e}^{\top}\mathbf{x} ,
\label{eq:global_en_relax}
\end{equation}
s.t. $(\mathbf{R}_0)$ and
\begin{equation}
(\mathbf{\widetilde{X}}_0)
\begin{aligned}
\left\{	\begin{array}{ll}
(\mathcal{C}_5) \hphantom{1} x_{i}^{l}+\sum_{j=1}^{M+1}x_{ij}^{f}+\sum_{j=1}^{M+1}x_{ij}^{c}=1, \forall i\in \mathbb{N},	\\
x_{i}^{l},x_{ij}^{f}, x_{ij}^{c} \in[0,1], \forall (i,j)\in \mathbb{N}\times\mathbb{M^*}.
\end{array}	\right.
\end{aligned}
\label{eq:17_offload_relax_con}
\end{equation}

We first will prove that the fully-relaxed problem $(\mathbf{\widetilde{P}}_0)$ is a convex optimization problem. 

\begin{theorem}
	\label{theo:convexity}
	The relaxed problem $(\mathbf{\widetilde{P}}_0)$ is a convex optimization problem.
\end{theorem}


{\emph{Proof:}} The detailed proof is presented in Appendix~\ref{sec:theo_convexity}.

The below solution to $(\mathbf{P}_0)$ obtained through solving $(\mathbf{\widetilde{P}}_0)$ is referred to as ``Relaxing Optimization Policy'' (ROP).
In ROP, we first solve the fully-relaxed optimization problem $(\mathbf{\widetilde{P}}_0)$. The convex optimization problem with constraints can be solved efficiently by the \emph{interior-point method}~\cite{Boyd2004Convex}, which is implemented in many popular solvers such as CPLEX, MOSEK, and the \text{fmincon()} function in MATLAB. Then, the real offloading decision solution of $(\mathbf{\widetilde{P}}_0)$ is converted to the closest integer decision for the problem $(\mathbf{P}_0)$. 
Due to the interdependence between the resource allocation and the offloading decision, the resource allocation solution of the relaxed problem has to be re-visited after rounding the integer decisions. Specifically, after fixing the offloading strategy  with converted integer decisions, we again solve the resource allocation of problem $(\mathbf{P}_0)$ to find feasible solutions.

Although this approximation method (ROP) can quickly find a solution, the solution is suboptimal. 
In the following sections, we introduce two effective methods to find the optimal solution of $(\mathbf{P}_0)$.

\subsection{Improved Branch and Bound Algorithm}

The conventional BB works by searching through a tree, in which every node of it represents a subproblem after fixing a binary variable. The relaxed version of the subproblem at that node, which is equivalent to a partly-relaxed problem of the original one, can be solved to evaluate the potential of that node before branching and visiting the left or right children nodes. As such, in the standard BB, the size of subproblems at children nodes is $1$ unit less than that of the father node in term of the number of variables. In other words, the complexity of the subproblems is reduced slowly in the conventional BB.

In this section, we introduce an improved branch and bound algorithm, namely IBBA, which efficiently solves the MINLP $(\mathbf{P}_0)$ by leveraging the unique characteristics of its binary decision variables to reduce the complexity. The IBBA, summarized in Algorithm~\ref{IBBA_Algorithm_code}, has the following features:

\begin{itemize}
	
\item \label{Branching_Task} \textbf{Branching task} dictates that a task can be executed at only one place, i.e, at the mobile device, one of the fog nodes, or the cloud server. 
Thus, for the offloading decisions $\mathbf{x}_i$ of task $i$ there is only one variable that is equal to $1$, and all others are equal to $0$. Thus, at every node in the search tree of IBBA, we choose to branch the decisions of a task, forming a $(2(M+1)+1)$-tree with height $N$. 

\item \label{Simplifying_problem} \textbf{Simplifying problem} dictates that when a task is executed at the mobile device, a fog node, or the cloud server, 
all other fog nodes do not need to allocate resources toward that task. Thus, when $x_{ij}^f = 0$ or $x_{ij}^c = 0$, we can eliminate all sub-expressions of the forms $x_{ij}^fA$ and $x_{ij}^cB$, these decision variables, and related resource allocation variables $r_{ij}^u$, $r_{ij}^d$, and $r_{ij}^f$ in~$(\mathbf{P}_0)$. Consequently, we have subproblems, namely $(\mathbf{RS})$, with a less number of variables.

\item \label{Preserving_Convexity} \textbf{Preserving convexity} dictates that the relaxed versions of subproblems $(\mathbf{RS})$ are convex optimization problems. In particular, based on Theorem~\ref{theo:convexity}, it can be observed that if we fix some binary variables in~$(\mathbf{P}_0)$ and set all other binary variables to be real ones, the corresponding relaxed subproblems, also called partly-relaxed problems, are always convex. 
\end{itemize}

The superiority of IBBA over the standard BB method in terms of complexity reduction is analyzed and presented in Section~\ref{sec:Complexity_Analysis}.




\emph{IBBA's Optimal Solution Selection}: The joint offloading and resource allocation problem may have more than one optimal solutions in which the numbers of tasks processed at the mobile device, edge nodes and the cloud server are different. While some network operators prefer the optimal solutions with more tasks processed at fog nodes, the others may not. For example, if the cost of cloud's computation resource is lower than that of fog nodes', then the network operators may prefer offloading to the cloud server. Otherwise, the fog computing is more preferable since it can reduce the backhaul throughput between fog nodes and the cloud server. However, general solvers for MIP, e.g., CPLEX and MOSEK, do not allow us to select an optimal solution. In the IBBA, the selection of an optimal solution can be realized by leveraging the unique structure of the joint offloading and resource allocation problem.

Without loss of generality, we develop an optimal solution selection policy, namely LFC (L, F and C, respectively, stands for local, fog and cloud processing), so that the final solution is the optimal one with tasks preferably processed at mobile device, then at fog nodes, and at the cloud server. In other words, if a problem has more than one optimal solutions, then the solution with more tasks processed locally at the mobile devices will be chosen as the final solution. Otherwise, if these optimal solutions have the same number of tasks processed locally, then the optimal solution with more tasks processed at fog nodes will be chosen as the final solution (and so on). 
Recall that the IBBA is a search tree-based algorithm, and the first optimal solution during the search is returned as the final solution. As mentioned above, \textbf{branching task} at a node will create a list of children nodes, each is equivalent to a fixed offloading decision of that computational task. 
Therefore, to enable the LFC policy, the list of children nodes will be visited in the order of offloading decisions, i.e., processing at mobile device, at fog nodes, and at cloud server.

\begin{figure}[!]
	\centering
	\includegraphics[width=0.95\linewidth]{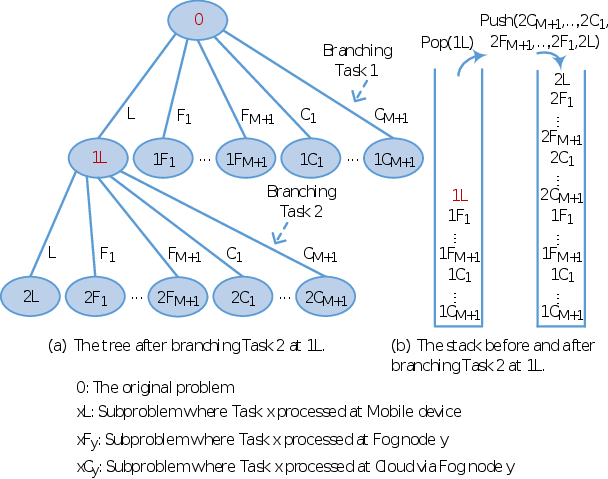}
	\caption{Search Tree for IBBA with Optimal Solution Selection.}
	
	\label{fig:SearchTree}
\end{figure}

Generally, we assume that computational tasks are chosen to branch in the order of $I_1,\dots, I_N$, and the processors (i.e., mobile device, fog nodes and cloud server) are chosen to serve these tasks in the order of $L, F_1, \dots, F_{(M+1)}, C_1, \dots, C_{(M+1)}$. Here, $I_i$ stands for task $I_i$, $L$ is a mobile device, $F_j$ is fog node $j$, and $C_j$ is for the case in which fog node $j$ forwards the computational task to the cloud server. The search tree in the IBBA method with this optimal solution selection policy, LFC, depicted in Fig.~\ref{fig:SearchTree}, is described as follows.

\begin{itemize}
	
	\item The \textit{deep-first search} (DFS) algorithm is used to travel the search tree of IBBA, and a \textit{stack} is used to store the subproblems $(\mathbf{RS})$, which are generated during traveling the tree. Here, the stack is a popular data structure for adding and removing subproblems at one end called top of the stack as in Fig.~\ref{fig:SearchTree}(b).
	
	\item Tasks are chosen to branch in the order of $I_1,\dots, I_N$. Tasks are called \textit{active} tasks if their offloading decisions have not been fixed on the search tree. 
	
	\item When branching a task, i.e., deciding where the task to be processed, the current subproblem on the top of the stack is deleted and the next subproblems $(\mathbf{RS})$ are generated and pushed in the stack in the order of processors $C_{(M+1)}, \dots, C_1, F_{(M+1)}, \dots, F_1, L$.
\end{itemize}

In Fig.~\ref{fig:SearchTree}(a), the DFS search algorithm scans the tree from left to right, and thus the most left branch optimal solution will be found first. Fig.~\ref{fig:SearchTree}(b) shows the status of the stack, which is a data structure to store the subproblems, before and after processing subproblem 1L. Here, the subproblem 1L on the top of the stack is deleted and the consequent subproblems are generated and pushed in the stack. We can see that the stack is suitable to the DFS search algorithm since the most left subproblem is always on the top of the stack. 
Consequently, the final solution is the optimal one with much tasks processed at mobile device, then at fog nodes, and at the cloud server.


In practice, we can sort tasks using different priority orders (e.g., their application types, or tasks' resource demand). For example, the higher-demand tasks are chosen to branch before the lower-demand tasks. 
Besides, each task can define its own order of processors. 
Thus, when branching a task at a node, the order of subproblems being pushed into the stack can be different according to its specific order of processors. 

In the IBBA, \textbf{Branching task} defines an efficient partition of the search space, and \textbf{Simplifying problem} only eliminates the cases which can not lead to a optimal solution. Note that the optimal solution selection schemes (i.e., LFC and LCF) do not reduce the search space. Thus, the IBBA's optimality is preserved, i.e., the same as that under the standard BB method. The detail of IBBA is summarized in Algorithm~\ref{IBBA_Algorithm_code}.  


%

{\renewcommand{\baselinestretch}{1}
\begin{algorithm}
	\DontPrintSemicolon
	\SetKwInput{Left}{left}\SetKwData{This}{this}\SetKwData{Up}{up}
	\SetKwInOut{Input}{Input}\SetKwInOut{Output}{Output}
	\Input{Set of tasks $\{I_{i}\left(D_{i}^{i},D_{i}^{o},C_{i},t_{i}^{r}\right)\}$\\ Set of ($M+1$) nodes $\{Node_j(R_{j}^{u},R_{j}^{d},R_{j}^{f})\}$\\ Cloud server $w^{c}$, $f^c$}
	\Output{Optimal solution and value of problem $(\mathbf{P}_0)$ }
	\BlankLine
	\Begin{
		$s \gets \emptyset$ \Comment{Initialize empty solution}\;
		$minE \gets +\infty$ \Comment{Initialize consumed energy +$\infty$}\;
		$t.empty()$ \Comment{Make stack empty}\;
		$t.push$($(\mathbf{P}_0)$) \Comment{Put $(\mathbf{P}_0)$ into stack}\;
		\While{$t.isNotEmpty()$}{
			$p \gets t.pop()$\Comment{Get subproblem from top of stack}\;
			$subs, subminE \gets$ \underline{Solve} relaxed problem of $p$ then return its optimal solution and value\;
			\If{$subminE > minE$ {\bf or} $p$ \textnormal{is infeasible}}{
				\underline{Prune} $p$ \Comment{Delete subproblem $p$}\;
			}
			\If{$subminE < minE$}{
				\If{$subs$ \textnormal{satisfies all integer constraints of} $\{\mathbf{x}_i\}$}{
					$s \gets subs$ \Comment{Update solution}\;
					$minE \gets subminE$\Comment{Update optimal result}\;
					\underline{Prune} $p$ \Comment{Delete subproblem $p$}\;
				}\Else{
					$children \gets$ \underline{Branch} $p$ by fixing the decisions of an \textit{active} task in the order of $\{I_i\}$ based on \textbf{Branching task} property.\;
					Sort $children$ in the order of increasing prority of processors.\;
					\For{{\bf{each}} $child$ {\bf in} $children$}{
						\underline{Simplify} $child$ based on \textbf{Simplifying problem} property.\;
						$t.push(child)$ \Comment{Put subproblem into stack}\; 
					}
				}
			}
		}
		\textbf{Return} $s$ and $minE$\;
	}
	\caption{IBBA Algorithm\label{IBBA_Algorithm_code}}
\end{algorithm}}

\subsection{Feasibility-Finding Benders Decomposition}



Although the IBBA has less and smaller size intermediate subproblems than the conventional BB algorithm, as analyzed in Section~\ref{sec:Complexity_Analysis}, the size of the intermediate subproblems reduces slowly. Moreover, the IBBA is not a distributed algorithm in essence. In this section, we design a distributed algorithm that decomposes the original problem into low-complexity subproblems that can be solved parallelly. Note that the dual decomposition method that is often instrumental to convex problem is not applicable to the underlying MINLP problem of which decisions for a task to be processed locally, at a fog, or at the cloud couple. 

The Benders method \cite{Geoffrion1972} transforms the original problem into a master problem and subproblems for both integer and continuous variables. These two simpler problems are solved iteratively so that Benders cuts (also called Benders cutting-planes) can be applied to both the
subproblems and the master problem.  
Note that the Benders decomposition method used in \cite{yu2018green} is not efficient for large scale problems. First, in \cite{yu2018green}, only one Benders cut is created in each iteration and the master problem may have to try most of its solutions. 
Second, this linearization method with the dual multipliers faces the zig-zagging issue~\cite{fischetti2011relax,fischetti2016benders} which increases the computation time.

\begin{figure}[!]
	\centering
	\begin{subfigure}[b]{0.4\textwidth}
		\centering
		\includegraphics[scale=0.48]{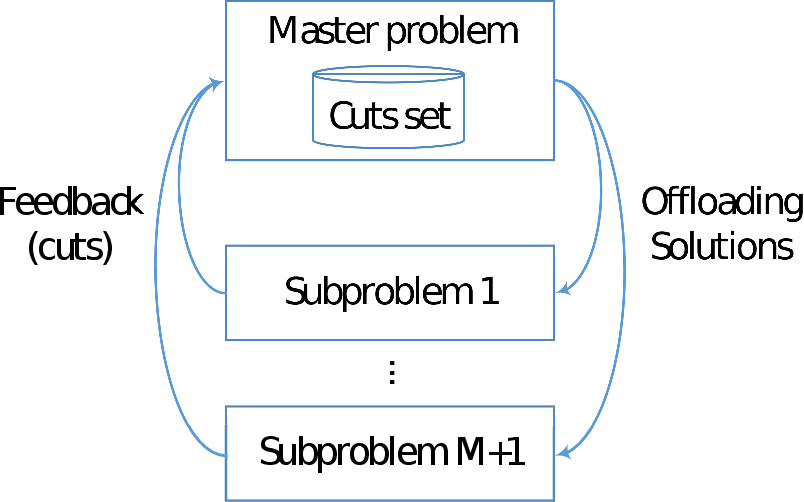}
		\caption{FFBD Model.}
		\label{sub.fig:ffbd-model}
	\end{subfigure}
	
	\begin{subfigure}[b]{0.4\textwidth}
		\centering
		\includegraphics[scale=0.60]{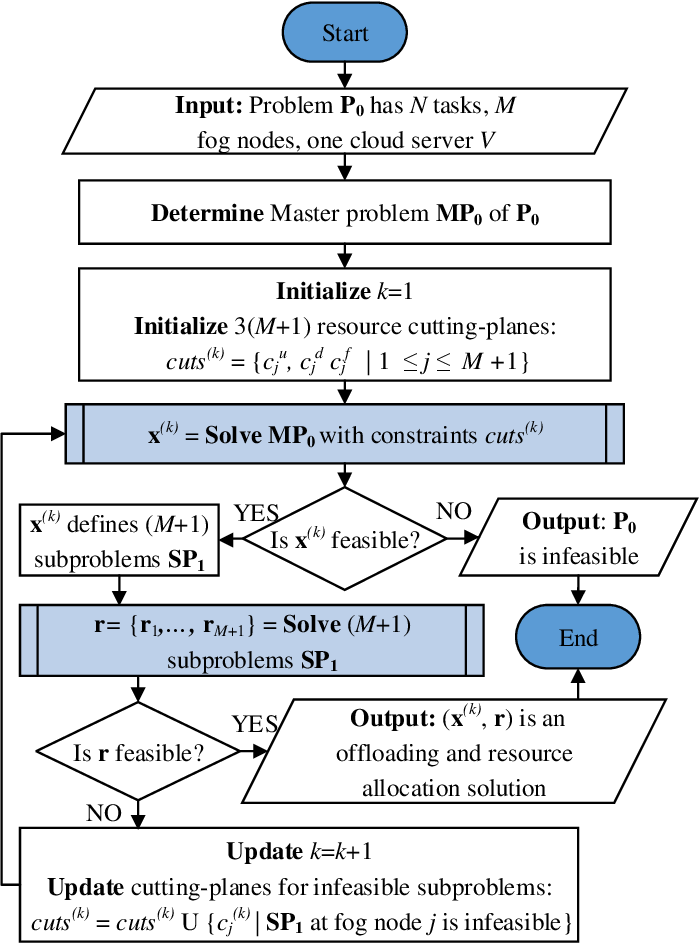}
		\caption{FFBD Procedure.}
		\label{sub.fig:ffbd-procedure}
	\end{subfigure}
	\caption{Feasibility-Finding Benders Decomposition.}
\label{fig:BD_Model}
\end{figure}

To tackle these issues, we develop a distributed algorithm named Feasibility-Finding Benders decomposition (FFBD) as illustrated in Fig.~\ref{fig:BD_Model}. The key idea of FFBD is the generation of \textit{Benders cuts} which exclude superfluous solutions, by assessing the communication and computation resources of fog nodes to satisfy tasks' requirements. Specifically, the Benders cuts are created simultaneously by solving the resource allocation problems at fog nodes. This completely differs from the approach presented in \cite{yu2018green} in which only one Benders cut is created by solving the dual problem in each iteration. 

%
%

Specifically, we first decompose $(\mathbf{P}_0)$ into a master problem $(\mathbf{MP}_0)$ for the offloading decision and a subproblem $(\mathbf{SP}_0)$ for the resource allocation. Then, the FFBD algorithm finds the optimal solution of $(\mathbf{P}_0)$ by iteratively solving $(\mathbf{MP}_0)$ and $(\mathbf{SP}_0)$ at either the cloud server or a fog node.
\begin{equation}
(\mathbf{MP}_0) \hphantom{10}	 \mathbf{x}^{(k)}=\underset{\mathbf{x} \in \mathbf{X}_0}{\argmin} \{\mathbf{e}^\top \mathbf{x} \}, 
\label{eq:master_problem}
\end{equation}
s.t. $cuts^{(k)},$
and
\begin{equation}
(\mathbf{SP}_0) \hphantom{10}	 \underset{\mathbf{r} \in \mathbf{R}_0}{\min}\{0\} ,
\label{eq:subproblem}
\end{equation}
where the set of Benders cutting-planes $cuts^{(k)}$ are restrictions on integer offloading solution $\mathbf{x}^{(k)}$ of $(\mathbf{MP}_0)$ at iteration $(k)$, and $\{0\}$ is the zero constant function.

From Eqs.~(\ref{eq:master_problem})~and~(\ref{eq:subproblem}), the cost function of $(\mathbf{P}_0)$ is identical with that of $(\mathbf{MP}_0)$. $(\mathbf{SP}_0)$ only verifies if integer offloading solution $\mathbf{x}^{(k)}$ of $(\mathbf{MP}_0)$ leads to a feasible resource allocation solution $\mathbf{r}$. Theorem~\ref{theo:ffbdstop} shows that the iteration can stop when a feasible solution $(\mathbf{x}, \mathbf{r})$ is found or $(\mathbf{MP}_0)$ is infeasible.
The convergence of the FFBD to an optimal solution in a finite number of iterations is analyzed in Section~\ref{sub:Cutting_Plane_Generation}.





\subsubsection{Distributed Subproblems}
At iteration $(k)$, by solving the integer programming problem $(\mathbf{MP}_0)$ the offloading decision variables $\mathbf{x}^{(k)}$ are found. The solution $\mathbf{x}^{(k)}$ determines if a task is processed at either mobile device, one of fog nodes, or the cloud server. Thus, $(\mathbf{SP}_0)$ is equivalently divided into a set of $(M+1)$ independent resource allocation subproblems (corresponding to $(M+1)$ nodes including the $M$ fog nodes and the cloud $V$). 

Let $\mathbf{x}_j=(\mathbf{x}_j^f,\mathbf{x}_j^c)$ be variables defining the offloading decisions of $N$ tasks onto fog node $j$ or the cloud server (via fog node $j$), where $\mathbf{x}_j^{f} = (x_{1j}^{f}, \dots, x_{Nj}^{f})$ and $\mathbf{x}_j^{c} = (x_{1j}^{c}, \dots, x_{Nj}^{c})$. Here, $\mathbf{x}_j$ is a part of $\mathbf{x}= \bigcup_{j =1}^{(M+1)} \mathbf{x}_j$.


Without loss of generality, we assume $\mathbb{N}_j^t$ and $\mathbb{N}_j^s$, respectively, be the sets of tasks to be processed at fog node $j$ and at the cloud server (via fog node $j$). Here, $\mathbb{N}_j^t$ and $\mathbb{N}_j^s$ are respectively determined by two offloading decision variables $\mathbf{x}_j^{f(k)}$ and $\mathbf{x}_j^{c(k)}$ in $\mathbf{x}^{(k)}$ at iteration $k$th. We can write $\mathbb{N}_j^t = \{1,\dots,t\}$, $\mathbb{N}_j^s = \{t+1,\dots,t+s\}$, and $\mathbb{N}_j^{t+s} =\mathbb{N}_j^t \cup \mathbb{N}_j^s= \{1,\dots,t+s\}$ is defined by $\mathbf{x}_j^{(k)} = (\mathbf{x}_j^{f(k)},\mathbf{x}_j^{c(k)})$. Variable $\mathbf{r}_j= (\mathbf{r}_{1j}, \dots, \mathbf{r}_{(t+s)j})$ is resources allocation of fog node $j$ towards its assigned set of tasks $\mathbb{N}_j^{t+s}$. Note that the fog node $j$ does not need to allocate the resources towards other tasks except $\mathbb{N}_j^{t+s}$. The resource allocation problem at fog node $j$ can be then written as
\begin{equation}
(\mathbf{SP}_1) \hphantom{10}	 
\underset{\mathbf{r}_j \in \mathbf{R}_j}{\min}\{0\} ,
\label{eq:sub_constrant_goal}
\end{equation}
where
\begin{equation}
\begin{aligned}
(\mathbf{R}_j)
\left\{	\begin{array}{ll}
(\mathcal{C}_{1j}) \hphantom{1} T_{i} \leq t_{i}^{r}, \forall i\in \mathbb{N}_j^{t+s},	\\
(\mathcal{C}_{2j}) \hphantom{1} \sum_{i \in \mathbb{N}_j^{t}}r_{ij}^{f} \leq R_{j}^{f}, \\
(\mathcal{C}_{3j}) \hphantom{1} \sum_{i \in \mathbb{N}_j^{t+s}}r_{ij}^{u} \leq R_{j}^{u}, \\
(\mathcal{C}_{4j}) \hphantom{1} \sum_{i \in \mathbb{N}_j^{t+s}}r_{ij}^{d} \leq R_{j}^{d},\\
r_{ij}^{f}, r_{ij}^{u}, r_{ij}^{d} \geq 0,\forall i \in \mathbb{N}_j^{t+s}, \\
r_{ij}^{f} = 0,\forall i \in \mathbb{N}_j^{s}.
\end{array}	\right.
\end{aligned}
\label{eq:resource_and_delay_constraints}
\end{equation}
As in Eq.~(\ref{eq:resource_delay_con}), $(\mathcal{C}_{1j})$ is the delay requirement of tasks, $(\mathcal{C}_{2j}), (\mathcal{C}_{3j})~\text{and}~ (\mathcal{C}_{4j})$ are resource constraints at fog node $j$.

Thus, instead of solving $(\mathbf{SP}_0)$, all subproblems $(\mathbf{SP}_1)$ can be solved distributedly among fog nodes in cooperation with the cloud server for $(\mathbf{MP}_0)$. Besides, these subproblems $(\mathbf{SP}_1)$ can also be solved in parallel at all fog nodes. Fig.~\ref{fig:BD_Model} shows the model of the distributed FFBD method.

\begin{theorem}
	\label{theo:ffbdstop}
	At any iteration $(k)$, if a feasible solution $(\mathbf{x})$ of $(\mathbf{MP}_0)$ leads to a feasible solution $(\mathbf{r})$ of $(\mathbf{SP}_0)$. Then, $(\mathbf{x},\mathbf{r})$ is the optimal solution of the original problem $(\mathbf{P}_0)$.
	
	At any iteration $(k)$, if the master problem~$(\mathbf{MP}_0)$ is infeasible, then the original problem $(\mathbf{P}_0)$ is infeasible.
\end{theorem}


{\emph{Proof:}} The detailed proof is presented in Appendix~\ref{sec:theo_ffbdstop}.



The solution of $(\mathbf{SP}_1)$ can be found by solving its equivalent problem $(\mathbf{SP}_2)$, which is always feasible,  with additional slack variables $\mathbf{z}$ using any solver. 
\begin{equation}
(\mathbf{SP}_2) \phantom{10}	 \underset{\mathbf{r}_j \in \mathbf{RZ}_j}{\min} \left(z_1 + z_2 + z_3\right) ,
\label{eq:sub_constrant_goal_relax_var}
\end{equation}
where
\begin{equation}
\begin{aligned}
(\mathbf{RZ}_j)
\left\{	\begin{array}{ll}
T_{i} \leq t_{i}^{r}, \forall i\in \mathbb{N}_j^{t+s},	\\
\sum_{i \in \mathbb{N}_j^{t}}r_{ij}^{f} - z_1 \leq R_{j}^{f}, \\
\sum_{i \in \mathbb{N}_j^{t+s}}r_{ij}^{u} - z_2 \leq R_{j}^{u}, \\
\sum_{i \in \mathbb{N}_j^{t+s}}r_{ij}^{d} - z_3 \leq R_{j}^{d},\\
r_{ij}^{f}, r_{ij}^{u}, r_{ij}^{d} \geq 0,\forall i \in \mathbb{N}_j^{t+s}, \\
r_{ij}^{f} = 0,\forall i \in \mathbb{N}_j^{s}, \\
z_1, z_2, z_3 \geq 0.
\end{array}	\right.
\end{aligned}
\label{eq:resource_and_delay_con}
\end{equation}

If $(\mathbf{SP}_2)$ is feasible and its cost function is zero, then $(\mathbf{SP}_1)$ is feasible. Otherwise, $(\mathbf{SP}_1)$ is infeasible.

At iteration $(k)$, if $(\mathbf{SP}_1)$ is feasible at every fog nodes, then $\mathbf{x}^{(k)}$ and $\mathbf{r}=(\mathbf{r}_{1}, \dots, \mathbf{r}_{(M+1)})$ are optimal solution of $(\mathbf{P}_0)$. Otherwise, if $(\mathbf{SP}_1)$ is infeasible at fog node $j$, a new cutting-plane $c_j^{(k)}$ will be added to the cut set of $(\mathbf{MP}_0)$ for the next iteration: $cuts^{(k+1)}=cuts^{(k)} \cup c_j^{(j)}$. The details of cutting-planes generation are in Section~\ref{sub:Cutting_Plane_Generation}.



\subsubsection{Fast Feasibility and Infeasibility Detection}
\label{sec:fast_feasible_infeasible}


Normally, the FFBD repeatedly solves $(\mathbf{MP}_0)$ and $M$ independent subproblems of the form $(\mathbf{SP}_1)$ using solvers, then update the cutting-plane set $cuts^{(k+1)}=cuts^{(k)} \cup c_j^{(j)}$. 
The closer to the optimal binary offloading decisions, the less number of iterations the master problem $(\mathbf{MP}_0)$ needs to be solved.  
Moreover, in many cases, we can quickly determine if $(\mathbf{SP}_1)$ is feasible or not without using any solver. Consequently, the computation time is reduced. The theoretical analysis below can be used to improve the efficiency of the FFBD algorithm.

From Eqs.~(\ref{eq:fog_delay}),~(\ref{eq:cloud_delay}),~and~(\ref{eq:direct_cloud_delay}), the delay constraint $(\mathcal{C}_{1j})~T_{i} \leq t_i^r$ in $(\mathbf{R}_j)$ of $(\mathbf{SP}_1)$ can be rewritten as
\begin{equation}
\begin{cases}

\left(\frac{D_{i}^{i}}{r_{ij}^{u}}+\frac{D_{i}^{o}}{r_{ij}^{d}}+\frac{C_{i}}{r_{ij}^{f}}\right) \leq t_i^r, & \forall i \in \mathbb{N}_j^t \\

\left(\frac{D_{i}^{i}}{r_{ij}^{u}}+\frac{D_{i}^{o}}{r_{ij}^{d}}\right) \leq t_i^r - 
\left(\frac{(D_{i}^{i}+D_{i}^{o})}{w^{c}}+\frac{C_{i}}{f^{c}}\right), & \forall i \in \mathbb{N}_j^s.
\end{cases}
\label{eq:delay_contraint1}
\end{equation}

Remarkably, the component $\left(\frac{(D_{i}^{i}+D_{i}^{o})}{w^{c}}+\frac{C_{i}}{f^{c}}\right)$ is a constant.
If $\exists i \in \mathbb{N}_j^s, t_i^{r} - \left(\frac{(D_{i}^{i}+D_{i}^{o})}{w^{c}}+\frac{C_{i}}{f^{c}}\right) \leq 0$, then processing task~$I_i$ at the cloud server does not satisfy its delay requirement $T_{i} \leq t_i^r$. In other words, $(\mathbf{SP}_1)$ is infeasible. A Benders cut to prevent offloading task $I_i$ to the cloud server can be directly created for this case.
Otherwise, if $t_i^{r} - \left(\frac{(D_{i}^{i}+D_{i}^{o})}{w^{c}}+\frac{C_{i}}{f^{c}}\right) > 0, \forall i \in \mathbb{N}_j^s$, then we define the relative size $(D_{i}^{i'}, D_{i}^{o'}, C_{i}^{'})$ of task $I_i$ as below. 
\begin{equation}
\begin{cases}
\left(\frac{D_{i}^{i}}{t_i^r}, \frac{D_{i}^{o}}{t_i^r}, \frac{C_{i}}{t_i^r}\right), & \forall i \in \mathbb{N}_j^t \\
\left(\frac{D_{i}^{i}}{\left(t_i^r - 
	\frac{(D_{i}^{i}+D_{i}^{o})}{w^{c}}-\frac{C_{i}}{f^{c}}\right)}, \frac{D_{i}^{o}}{\left(t_i^r - 
	\frac{(D_{i}^{i}+D_{i}^{o})}{w^{c}}-\frac{C_{i}}{f^{c}}\right)}, 0\right), & \forall i \in \mathbb{N}_j^s.
\end{cases}
\label{eq:define_task}
\end{equation}

Let $\beta_i = \left(
\frac{D_{i}^{i'}}{r_{ij}^{u}}+
\frac{D_{i}^{o'}}{r_{ij}^{d}}+
\frac{C_{i}^{'}}{r_{ij}^{f}}
\right)$ be the satisfaction rate of task $I_i$. The delay constraint in Eq.~(\ref{eq:delay_contraint1}) becomes
\begin{equation}
\begin{aligned}
\beta_i =
\left(
\frac{D_{i}^{i'}}{r_{ij}^{u}}+
\frac{D_{i}^{o'}}{r_{ij}^{d}}+
\frac{C_{i}^{'}}{r_{ij}^{f}}
\right) \leq 1, \forall i \in \mathbb{N}_j^{t+s}.
\end{aligned}
\label{eq:delay_constraints_new1}
\end{equation}


The Theorems~\ref{theo:feasible}~and~\ref{theo:infeasible} below can quickly check the feasibility and infeasibility of $(\mathbf{SP}_1)$.

\begin{theorem}
	\label{theo:feasible}
	When fog node $j$ allocates all resources proportional to the input, output data sizes and CPU cycles of tasks, the equivalent delay components of these tasks are equal and defined as 
	$\beta_{bal}^u = \frac{\sum_{i \in \mathbb{N}_j^{t+s}}
		D_i^{i'}}{R_j^u}$, $\beta_{bal}^d = \frac{\sum_{i \in \mathbb{N}_j^{t+s}}
		D_i^{o'}}{R_j^d}$, and $\beta_{bal}^f = \frac{\sum_{i \in \mathbb{N}_j^{t+s}}
		C_i^{'}}{R_j^f}$. If $\beta_{bal} = \beta_{bal}^u + \beta_{bal}^d + \beta_{bal}^f \leq 1$, then the problem $(\mathbf{SP}_1)$ is feasible.
\end{theorem}

{\emph{Proof:}} The detailed proof is presented in Appendix~\ref{sec:theo_feasible}.

%


\begin{corollary}
	\label{coro:feasible}
	
	Let $\beta_{bal}^f = \frac{\sum_{i \in \mathbb{N}_j^{t}}
		C_i^{'}}{R_j^f}$,
	$\gamma_{bal}^{u} = \frac{\sum_{i \in \mathbb{N}_j^{t}}
		D_i^{i'}/\left(1-\beta_{bal}^f\right)+\sum_{i \in \mathbb{N}_j^{s}}
		D_i^{i'}}{R_j^u}$, and
	$\gamma_{bal}^{d} = \frac{\sum_{i \in \mathbb{N}_j^{t}}
		D_i^{o'}/\left(1-\beta_{bal}^f\right)+\sum_{i \in \mathbb{N}_j^{s}}
		D_i^{o'}}{R_j^d}$.
	
	If $\beta_{bal}^f \leq 1$ and $\gamma_{bal} = \gamma_{bal}^{u} + \beta_{bal}^{d} \leq 1$, then the problem $(\mathbf{SP}_1)$ is feasible.
\end{corollary}


The Corollary~\ref{coro:feasible} is derived from Theorem~\ref{theo:feasible}. Fog node $j$ first allocates the computation resource toward tasks in the set $\mathbb{N}_j^{t}$, which are processed at the fog node, then allocates both the uplink and downlink resources toward all tasks in $\mathbb{N}_j^{t+s}$ based on their remaining delay requirements. Noticeably, tasks in $\mathbb{N}_j^{t+s}$ are either processed at fog node $j$ or forwarded to the cloud server by this fog node. Then, we apply sequentially Theorem~\ref{theo:feasible} to $\beta_{bal}^f$ and $\gamma_{bal}$.

\begin{lemma}
	\label{lem:inequality}
	
	Assume variables $p_i \geq 0$, $q_i > 0$, $\forall i \in N$, satisfying conditions: $\sum_{i \in N} p_i = P$ and $\sum_{i \in N} q_i = Q$.
	We have $\underset{i \in N}{\max} \{\frac{p_i}{q_i}\} \geq \frac{P}{Q}$.
\end{lemma}

{\emph{Proof:}} The detailed proof is presented in Appendix~\ref{sec:lem_inequality}.


\begin{theorem}
	\label{theo:infeasible}
	If $\frac{\sum_{i \in \mathbb{N}_j^{t+s}}
		D_i^{i'}}{R_j^u} > 1$ or  $\frac{\sum_{i \in \mathbb{N}_j^{t+s}}
		D_i^{o'}}{R_j^d} > 1$ or  $\frac{\sum_{i \in \mathbb{N}_j^{t+s}}
		C_i^{'}}{R_j^f} > 1$, then the problem $(\mathbf{SP}_1)$ is infeasible.
\end{theorem}

{\emph{Proof:}} The detailed proof is presented in Appendix~\ref{sec:theo_infeasible}.




\emph{Fast Feasibility Detection:}
Theorem~\ref{theo:feasible} helps find a feasible solution of the subproblem at fog node $j$ with assigned tasks $\mathbb{N}_j^{t+s}$ so that using any solver is not necessary. Consequently, the computation time is reduced. Especially, for the case of the ratio of $D_i^{i'}$, $D_i^{o'}$ and $C_i^{'}$ approximately equal between all tasks $\mathbb{N}_j^{t+s}$, this theorem can find feasible solutions of subproblems.


Corollary~\ref{coro:feasible} is theoretically stronger than Theorem~\ref{theo:feasible} because it repeatedly applies the theorem to the computation resource then calculates the remaining delay requirements to allocate the uplink and downlink resources. Therefore, we can use Corollary~\ref{coro:feasible} instead of Theorem~\ref{theo:feasible} for feasibility detection.

\emph{Fast Infeasibility Detection:}
The cutting-planes based on Theorem~\ref{theo:infeasible} are useful for the large scale system (e.g., thousands of tasks and hundreds of fog nodes). For example, if a fog node can approximately support a maximum of $n$ tasks, these cutting-planes can avoid the generation of subproblems with more than $n$ assigned tasks.

\subsubsection{Cutting-Plane Generation}
\label{sub:Cutting_Plane_Generation}

In this paper, we introduce three types of cutting-planes which will be updated in $(\mathbf{MP}_0)$, namely ``Resource Cutting-Plane'', ``Subproblem Cutting-Plane'',  and ``Prefixed Decision Cutting-Plane''. Although the FFBD can find the optimal solution only by using the subproblem cutting-planes (as analyzed below), by using the resource and prefixed decision cutting-planes, the master problem can avoid the offloading decisions that violate the resources and delay constraints. That helps reduce the search space.

\emph{Resource Cutting-Plane:}

Recall that, $\mathbf{x}_j^f$ is the offloading decision variable vector that determines the subset of tasks $\mathbb{N}_j^{t} \subseteq \mathbb{N}$ being processed at fog node $j$, and $\mathbf{x}_j^c$ is the offloading decision variable vector that determines the subset of tasks $\mathbb{N}_j^{s} \subseteq \mathbb{N}$ being sent to the cloud by fog node $j$. Thus, vector $(\mathbf{x}_j^f,\mathbf{x}_j^c)$ determines the subset of tasks $\mathbb{N}_j^{t+s} \subseteq \mathbb{N}$.

Let $\mathbf{c}_j^{u(fog)}$, $\mathbf{c}_j^{d(fog)}$ and $\mathbf{c}_j^{f(fog)}$, respectively, be the coefficient vectors of $\mathbf{x}_j^f$ in the uplink, downlink and computation resource cutting-planes below. 
From Theorem~\ref{theo:infeasible}, we have $\mathbf{c}_j^{u(fog)} = (D_1^{i'}, \dots, D_N^{i'} )/R_j^u$, $\mathbf{c}_j^{d(fog)} = (D_1^{o'}, \dots, D_N^{o'})/R_j^d$ and $\mathbf{c}_j^{f(fog)} = (C_1^{'}, \dots, C_N^{'})/R_j^f$. Here, $(D_i^{i'}, D_i^{o'}, C_i^{'})$ is calculated as in Eq.~(\ref{eq:define_task}) for $i \in \mathbb{N}_j^t$.

Let $\mathbf{c}_j^{u(cloud)}$, $\mathbf{c}_j^{d(cloud)}$ and $\mathbf{c}_j^{f(cloud)}$, respectively, be the coefficient vectors of $\mathbf{x}_j^c$ in the uplink, downlink and computation resource cutting-planes below. 
From Theorem~\ref{theo:infeasible}, we have  $\mathbf{c}_j^{u(cloud)} = (D_1^{i'}, \dots, D_N^{i'})/R_j^u$, $\mathbf{c}_j^{d(cloud)} = (D_1^{o'}, \dots, D_N^{o'})/R_j^d$ and $\mathbf{c}_j^{f(cloud)} = (C_1^{'}, \dots, C_N^{'})/R_j^f$. Here, $(D_i^{i'}, D_i^{o'}, C_i^{'})$ is calculated as in Eq.~(\ref{eq:define_task}) for $i \in \mathbb{N}_j^s$.

From Theorem~\ref{theo:infeasible}, to avoid the generation of every subset $\mathbb{N}_j^{t+s} \subseteq \mathbb{N}$ that violates the uplink, downlink and computation resource constraints at edge node~$j$, we add three following Benders cuts into $cuts$ set of the Master problem $(\mathbf{MP}_0)$:
$$c_j^u=\{\mathbf{c}_j^{u(fog)\top} \mathbf{x}_j^f + \mathbf{c}_j^{u(cloud)\top}\mathbf{x}_j^{c} \leq 1\},$$
$$c_j^d=\{\mathbf{c}_j^{d(fog)\top} \mathbf{x}_j^f + \mathbf{c}_j^{d(cloud)\top}\mathbf{x}_j^{c} \leq 1\}, ~\text{and}$$
$$c_j^f=\{\mathbf{c}_j^{f(fog)\top} \mathbf{x}_j^f + \mathbf{c}_j^{f(cloud)\top}\mathbf{x}_j^{c} \leq 1\}.$$

From Eq.~(\ref{eq:define_task}), we have $\mathbf{c}_j^{f(cloud)} = \mathbf{0}$. Therefore, $c_j^f$ can be shorten as
$c_j^f=\{\mathbf{c}_j^{f(fog)\top} \mathbf{x}_j^f \leq 1\}.$

\emph{Subproblem Cutting-Plane:}
At iteration $(k)$, fog node $j$ is assigned a set of tasks $\mathbb{N}_j^{t+s} =\mathbb{N}_j^t \cup \mathbb{N}_j^s$, which is defined by offloading decision $\mathbf{x}_j^{(k)} = (\mathbf{x}_j^{f(k)},\mathbf{x}_j^{c(k)})$. 
If the resource allocation problem $(\mathbf{SP}_1)$ at fog node $j$ is infeasible, then any resource allocation problem at edge node $j$ with assigned tasks $\mathbb{N}_j \supseteq \mathbb{N}_j^{t+s}$ is infeasible. Thus, to eliminate all subproblems at edge node $j$ containing $\mathbb{N}_j^{t+s}$, a new Benders cut $c_j^{(k)}$ is added into $cuts$ set of the Master problem $(\mathbf{MP}_0)$ after iteration $(k)$:  $$c_j^{(k)}=\{\mathbf{x}_{j}^{f(k)\top} \mathbf{x}_{j}^{f} + \mathbf{x}_{j}^{c(k)\top} \mathbf{x}_{j}^{c} \leq t+s-1\}.$$ 


\emph{Prefixed Decision Cutting-Plane:}
If task $I_i$ satisfies $E_{i}^{l} < E_{ij}^{f}$ and $T_{i}^{l} \leq t_{i}^{r}$, then it can be pre-decided as local processing. As mentioned in \textit{Fast Feasibility and Infeasibility Detection}, if $\left(t_i^{r} - \frac{(D_{i}^{i}+D_{i}^{o})}{w^{c}}-\frac{C_{i}}{f^{c}}\right) \leq 0$, then task $I_i$ could not be processed at the cloud cluster. In these cases, the suitable cutting-planes can be created and added to set $cuts$ of $(\mathbf{MP}_0)$.


In each iteration of the FFBD, if a subproblem~$(\mathbf{SP}_1)$ is infeasible then a \textit{subproblem cutting-plane} is created. Each \textit{subproblem cutting-plane} is equivalent to a set of computational tasks, which are either processed or forwarded to the cloud server for execution by fog node $j$. 
Besides, due to the finite numbers of tasks and fog nodes, the number of \textit{subproblem cutting-planes} is finite. Consequently, the FFBD stops after has a finite number of iterations. Based on Theorem~\ref{theo:ffbdstop} we can conclude that the FFBD always returns the optimal solution after a limited number of iterations. This is equivalent to the conditions to converge to an optimal solution in the standard Benders decomposition~\cite{Benders1962Partitioning,Grothey1999note,Taskin2011Benders}.

{\renewcommand{\baselinestretch}{1}
\begin{algorithm}
	\DontPrintSemicolon
	\SetKwInput{Left}{left}\SetKwData{This}{this}\SetKwData{Up}{up}
	\SetKwInOut{Input}{Input}\SetKwInOut{Output}{Output}
	\Input{Set of tasks $\{I_{i}\left(D_{i}^{i},D_{i}^{o},C_{i},t_{i}^{r}\right)\}$\\ Set of ($M+1$) nodes $\{Node_j(R_{j}^{u},R_{j}^{d},R_{j}^{f})\}$\\ Cloud server $w^{c}$, $f^c$}
	\Output{Optimal solution $(\mathbf{x},\mathbf{r})$ of Problem $(\mathbf{P}_0)$ }
	\BlankLine
	\Begin{
		Initialize $k$ and $cuts^{(k)}$ as in \textbf{Initialization}.\;
		\While{\textnormal{solution} $(\mathbf{x},\mathbf{r})$ \textnormal{has not been found}}{
			
			$\mathbf{x} \gets$ \underline{Solve} $(\mathbf{MP}_0)$ with $cuts^{(k)}$ as in \textbf{Master Problem}. \Comment{$\mathbf{x}$ store solution $\mathbf{x}^{(k)}$ at iteration $k$}\;
			
			\If{$\mathbf{x}$\textnormal{ is feasible}}{
				Based on $\mathbf{x}$, create $(M+1)$ subproblems $(\mathbf{SP}_1)$ with asigned tasks $\mathbb{N}_1^{t+s}, \dots, \mathbb{N}_{M+1}^{t+s}$.\;
			}\Else{
				\textbf{Return} Problem $(\mathbf{P}_0)$ is infeasible.\;
			}
			
			\For{ $(j = 1;\ j \leq (M+1);\ j = j + 1)$}{
				$\mathbf{r}_j \gets $ \underline{Solve} $(\mathbf{SP}_1)$ at fog node $j$ with task set $\mathbb{N}_j^{t+s}$ as in \textbf{Subproblems}.\;
				
				\If{$\mathbf{r}_j$\textnormal{ is infeasible}}{
					
					Add a new Benders cut $c_j^{(k)}$ into $cuts^{(k+1)}$ as in \textbf{Subproblems}.
				}		
			}
			
			\If{$\mathbf{r} = (\mathbf{r}_1 \cup \mathbf{r}_2 \cup \dots \cup \mathbf{r}_{(M+1)})$\textnormal{ is feasible}}{
				Solution $(\mathbf{x},\mathbf{r})$  has been found.\;
			}
			$k \gets (k+1)$ \Comment{Increase iteration index}\;
		}
		\textbf{Return} $\mathbf{x}$ and $\mathbf{r}$\;
	}
	\caption{FFBD Algorithm
		\label{FFBD_Algorithm_code}}
\end{algorithm}}






\subsubsection{FFBD Procedure}


The operation of the distributed FFBD algorithm is summarized in Fig.~\ref{fig:BD_Model}. 
At the iteration $(k)$, the offloading decision solution $\mathbf{x}^{(k)}$ of $(\mathbf{MP}_0)$ determines where the tasks in $\mathbb{N}$ will be processed (i.e., the mobile device, fog nodes and the cloud server). 
Assume $\mathbb{N}_j^{s+t} \subseteq \mathbb{N}$ to be the set of tasks assigned to fog node $j$. Then every fog node $j$ independently solves its own resource allocation problem of the form $(\mathbf{SP}_1)$. 
The Feasibility-Finding Benders decomposition, Algorithm~\ref{FFBD_Algorithm_code}, is described as below.

\begin{itemize}
	
	\item \label{BD_initializaiton} \textbf{Initialization:}
	Set the iterator $k=1$.
	Then, initialize $cuts^{(k)}$ in $(\mathbf{MP}_0)$ with $3(M+1)$ resource cutting-planes as in Section~\ref{sub:Cutting_Plane_Generation}:  $cuts^{(k)}=\bigcup_{j=1}^{M+1} \{c_j^u, c_j^d, c_j^f\}.$
	
	Other Benders cuts (i.e., prefixed decision cutting-planes) as in Section~\ref{sub:Cutting_Plane_Generation}, are also added to the $cuts^{(k)}$ of $(\mathbf{MP}_0)$.
	
	
	
	
	
	\item \label{BD_master_problem} \textbf{Master Problem:} At iteration $(k)$, $(\mathbf{MP}_0)$ is solved to find $\mathbf{x}^{(k)} \in X_0$ satisfying $cuts^{(k)}$. Here, $\mathbf{x}^{(k)}$ defines $(M+1)$ subproblems of the form $(\mathbf{SP}_1)$.
	If $(\mathbf{MP}_0)$ is infeasible, then the FFBD is terminated with the infeasibility of $(\mathbf{P}_0)$. 
	If $(\mathbf{MP}_0)$ with its solution $\mathbf{x}^{(k)}$ leads to a feasible solution $\mathbf{r} = (\mathbf{r}_1 \cup \mathbf{r}_2 \cup \dots \cup \mathbf{r}_{(M+1)})$ of $(M+1)$ subproblems of the form $(\mathbf{SP}_1)$, then the FFBD is terminated, and  $(\mathbf{x}^{(k)}, \mathbf{r})$ is the optimal feasible solution of the original problem~$(\mathbf{P}_0)$.
	
	\item \label{BD_subproblem} \textbf{Subproblems:} At iteration $(k)$, a set of computational tasks $\mathbb{N}_j^{t+s} = \mathbb{N}_j^{t}\cup \mathbb{N}_j^{s}$ are assigned to fog node $j$, in which $\mathbb{N}_j^{t}$ and $\mathbb{N}_j^{s}$ are, respectively, processed and forwarded to the cloud for execution by fog node $j$. Then, fog node $j$ independently solves $(\mathbf{SP}_1)$ in order to allocate resource to its own assigned tasks.
	Before calling a solver,  Theorem~\ref{theo:feasible} is used to check its feasibility. If $(\mathbf{SP}_1)$ is infeasible, then a new subproblem cutting-plane $c_j^{(k)}$ as in Section~\ref{sub:Cutting_Plane_Generation} is created and added into $cuts^{(k+1)}$ of $(\mathbf{MP}_0)$ for the next iteration $(k+1)$:
	$cuts^{(k+1)}=cuts^{(k)}\cup\{c_j^{(k)}\}.$ 
\end{itemize}

\subsection{Implementation Protocol and Complexity Analysis}

\subsubsection{Implementation Protocol}

We first introduce a method to identify the user devices, then propose a protocol that defines the operation of the ROP, IBBA and FFBD methods. 

\emph{Device Identification:}
In order to cooperate horizontally and vertically, all devices, i.e., mobile devices, fog nodes and cloud server need to have a unique identification (ID), which can be determined by the MAC address or a temporary granted number. We assume that fog nodes and the cloud server have permanent IDs, e.g., MAC address. 
For the mobile devices without the permanent IDs, they can be periodically granted two-component temporary IDs of the form $(\emph{\mbox{FID}}_j,\emph{\mbox{NID}})$, in which $\emph{\mbox{FID}}_j$ is the permanent ID of fog node $j$, and $\emph{\mbox{NID}}$ is an integer number managed by fog node $j$. To make a unique temporary ID $(\emph{\mbox{FID}}_j,\emph{\mbox{NID}})$, fog node $j$ will allocate different $\emph{\mbox{NID}}$ in each period. Besides, each mobile device can hold at most one temporary ID $(\emph{\mbox{FID}}_j,\emph{\mbox{NID}})$ provided by at most one fog node. Due to the cooperation among fog nodes and the cloud server, the mobile devices can be managed using these temporary IDs. 

%
%
%
%

\begin{figure}[!]
	\centering
	\includegraphics[scale=0.50]{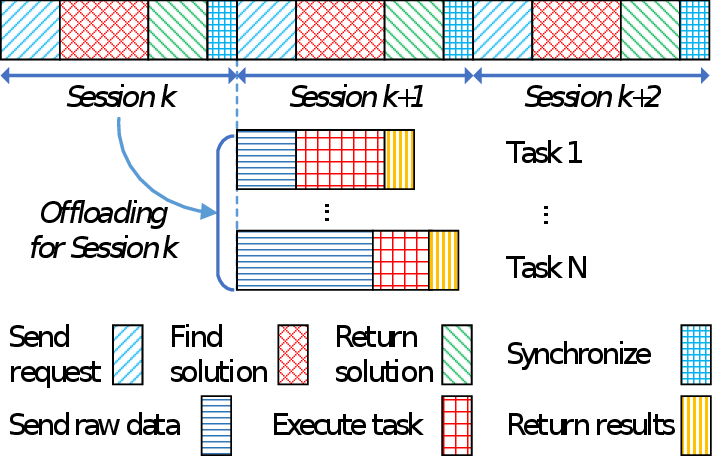}
	\caption{Protocol defining the operation of proposed methods.}
	\label{fig:Protocol-for-methods}
\end{figure}

\emph{Proposed Protocol:}
The timeline of the fog computing system is divided into \textit{session} including a fixed number of adjacent time slots. A \textit{sessions}, then, includes four stages, namely \textbf{Send request}, \textbf{Find solution}, \textbf{Return solution} and \textbf{Synchronize} as in Fig.~\ref{fig:Protocol-for-methods}. 
Here, we assume that the communication and computation resources are managed so that the operation of the protocol in the current session (i.e., \textit{session k}) and the execution of the tasks from previous sessions (i.e., \textit{session k-1, k-2}) can occur simultaneously. 
The following are the details of these stages.

\begin{itemize}
	
	\item \textbf{Send request:} At the begin of \textit{session k}, mobile devices send offloading requests containing a unique identification $(\emph{\mbox{FID}}_j, \emph{\mbox{NID}}, \emph{\mbox{tk}})$, where $\emph{\mbox{tk}}$ is a number to identify the task, and the task's information.
	
	\item \textbf{Find solution:} 
	At this stage, a fog node or the cloud server with powerful computation capacities will run either the ROP, IBBA or FFBD. The FFBD method can also be run in a parallel or distributed cooperative manner. Noticeably, the tasks of \textit{session k} will be offloaded then executed from the beginning of \textit{session k+1}, thus all fog nodes need to estimate their available resources from \textit{session k+1} to find the solution in \text{session k}.
	
	\item \textbf{Return solution:} The joint offloading and resource allocation solution found in the \textbf{Find solution} stage will be sent to mobile devices via fog nodes.
	
	\item \textbf{Synchronize:} This gap period is for synchronizing the fog computing system, preparing the offloading requests and estimating the available resources for the next session.
	
	
	
\end{itemize}


Since the timeline is divided into adjacent sessions as presented in Fig.~\ref{fig:Protocol-for-methods}, the offloading, executing and returning the results of tasks for \textit{session k} can spread over the next sessions, e.g., \textit{sessions k+1, k+2}.

\subsubsection{Complexity Analysis}
\label{sec:Complexity_Analysis}

In this section, we evaluate the complexity of the proposed methods in terms of problems' size, which is affected by both the numbers of tasks and  fog nodes.



\emph{Original Problem:}
With $(M+1)$ nodes (including $M$ fog nodes and the cloud $V$), the original problem $(\mathbf{P}_0)$ has $3(M+1)$ resource constraints as  $(\mathcal{C}_2), (\mathcal{C}_3)~\text{and}~ (\mathcal{C}_4)$ described in Eq.~(\ref{eq:resource_delay_con}). 
Besides, to formulate each task $I_i$ in $(\mathbf{P}_0)$, it needs $(2(M+1)+1)$ binary and $3(M+1)$ real variables, 
two delay and offloading constraints $(\mathcal{C}_1)$ as in Eq.~(\ref{eq:resource_delay_con}) and $(\mathcal{C}_5)$ as in Eq.~(\ref{eq:offload_qos_con}). 
Therefore, with $N$ tasks and $(M+1)$ nodes (including $M$ fog nodes and the cloud $V$), the original problem $(\mathbf{P}_0)$ has respectively $N(2(M+1)+1)$ integer and $\emph{\mbox{3N(M+1)}}$ real variables, and $(2N+3(M+1))$ constraints including $N$ for the offloading decisions, $N$ for the delay of the tasks and $3(M+1)$ for the resource requirements of the fog nodes as described in Eq.~(\ref{eq:resource_delay_con}) and Eq.~(\ref{eq:offload_qos_con}).
Consequently, the relaxed problem $(\mathbf{\widetilde{P}}_0)$ of $(\mathbf{P}_0)$ has totally $N(5(M+1)+1)$ real variables and $(2N+3(M+1))$ constraints.

\emph{ROP Method:}
For the ROP method, we solve the relaxed problem $(\mathbf{\widetilde{P}}_0)$ once, then convert the real offloading decision solution to the closest integer decision for the problem $(\mathbf{P}_0)$. To improve the performance of the ROP method, we fix the offloading strategy with the converted integer decisions, then solve the resource allocation problem $(\mathbf{P}_0)$. Thus, the ROP has low computation time. However, this approximation method is not a reliable method since the final offloading solution may be suboptimal and the constraints are not guaranteed to be met.

\emph{Standard BB Method:}
For the standard BB method, it works as a binary search tree, in which every node on the tree represents a subproblem after fixing a binary offloading variable. Thus, in the worst case, the standard BB method has to solve $(2^1 + 2^2 + \dots + 2^{N(2(M+1)+1)}) = (2^{N(2(M+1)+1)+1} - 2)$ intermediate relaxed problems with the size decreasing from  $(N(5(M+1)+1)-1)$ to $(3N(M+1))$ real variables. 

\emph{IBBA Method:}
For the IBBA method, it works as a $(2(M+1)+1)$-tree, in which every node on the tree represents a subproblem after deciding where a task is processed. Thus, in the worst case, the IBBA method has to solve $((2(M+1)+1)^1 + (2(M+1)+1)^2 + \dots + (2(M+1)+1)^{N}) = \frac{((2(M+1)+1)^{N+1} - (2(M+1)+1))}{(2(M+1))} \approx  (2(M+1)+1)^{N+1}/(2(M+1))$ intermediate relaxed problems with the size decreasing from  $((N(5(M+1)+1)-(2(M+1)+1) - 3(M))$ to $(3N)$ real variables. 
Noticeably, the IBBA method updates the current solution by traveling the branches of the search tree. If it finds a new solution which is better than the current solution, then the current solution will be updated by the new one. Otherwise, if the relaxed solution of the intermediate problem at a node is no better than the current solution, then the branch starting from that node will be pruned. Here, a branch is equivalent to a large number of intermediate problems. Therefore, most of the intermediate problems are eliminated.
We can see that the proposed IBBA method has to solve fewer intermediate relaxed problems with smaller size in comparing with the conventional BB method.



\emph{FFBD Method:}
For the FFBD method, the master problem $(\mathbf{MP}_0)$ and $M+1$ subproblems of the form $(\mathbf{SP}_1)$ are iteratively solved. At iteration $k$, $(\mathbf{MP}_0)$ is an integer problem with $N(2(M+1)+1)$ binary offloading variables and at most $(N+3(M+1) + k(M+1)$ constraints including $N$ for the offloading decision constraints as $(\mathcal{C}_5)$ described in Eq.~(\ref{eq:offload_qos_con}), $3(M+1)$ for resource cutting-planes as in the \textbf{Initialization} step, and at most $k(M+1)$ for the subproblem cutting-planes from solving $(M+1)$ subproblems $k$ times as in \textbf{Subproblem} step. Besides, each subproblem $(\mathbf{SP}_1)$ is assigned an average of $N/(M+1)$ tasks. Thus, it has approximate $3N/(M+1)$ resources allocation variables and $(N/(M+1)+3)$ constraints including $N/(M+1)$ for the delay of $N/(M+1)$ tasks as $(\mathcal{C}_{1j})$ in Eq.~(\ref{eq:resource_and_delay_constraints}) and $3$ constraints for the resources requirements at the fog node as $(\mathcal{C}_{2j})$, $(\mathcal{C}_{3j})$  and $(\mathcal{C}_{4j})$ described in Eq.~(\ref{eq:resource_and_delay_constraints}). In the worst case, $(\mathbf{SP}_1)$ is assigned all $N$ tasks, thus, has at most $3N$ resources allocation variables and $(N+3)$ constraints. However, if this big subproblem violates the resources constraints at the fog node according to Theorem~\ref{theo:infeasible}, it could not be created due to the resources cutting-planes generation in the \textbf{Initialization} step.

We can see that the size of the MP and subproblems in the FFBD is a linear function of the number of tasks, whereas the size of the intermediate problems in the IBBA is exponential w.r.t. the number of tasks. 

\section{Performance Evaluation} 
\label{sec:performanceevaluation}

\subsection{Offloading Analysis}
Before conducting experiments, we analyze when mobile users can benefit from offloading. Let $\alpha_i$ be the ratio between the number of required CPU cycles $C_i$ and input data size $D_i^i$. We have $C_i = \alpha_i \times D_i^i$. Then, the local consumed energy is $E_{i}^{l}=v_{i}C_{i}= \alpha_i v_{i} D_i^i$.

A mobile user is said to benefit from offloading if its total energy consumption from task offloading is lower than being locally processed. 
Thus, for task $I_i$, offloading will benefit if $E_i^l > E_i^f$. In other words, we have $\alpha_i v_{i} D_i^i > e_{ij}^uD_i^i + e_{ij}^dD_i^o$. 
Let $\alpha_i^*$ be the task complexity ratio at which $E_i^l = E_i^f$. We have:
\begin{equation}
\alpha_i^* = \frac{e_{ij}^uD_i^i + e_{ij}^dD_i^o}{v_i D_i^i}.
\label{eq:complexity_rate}
\end{equation}

Thus, task $I_i$ is likely to be offloaded if $E_i^l > E_i^f$ or $\alpha_i > \alpha_i^*$. The task complexity $\alpha_i$ is especially important in evaluating offloaded tasks as well as analyzing the performance of the whole system. 

\subsection{Experiment Setup} 
We use the configuration of a Nokia N900 mobile device as in \cite{miettinen2010energy} and set the number of devices as $N = 10$. Each mobile device has CPU rate $f_i^l=0.5$ Giga cycles/s and the unit processing energy consumption $v_i=\frac{1000}{730}$ J/Giga cycle (Energy characteristics of local computing for Nokia N900/500~MHz in Table~1~in~\cite{miettinen2010energy}).
In the IoT ecosystem, offloading demand applications often have different characteristics in term of tasks' data size and complexity. These applications share the same communication and computation resources of fog nodes and the cloud server. Therefore, it is reasonable to choose randomly data size
and complexity.
We denote $U(a,b)$ as the discrete uniform distribution between $a$ and $b$. Here, we assume that the input and output data sizes following uniform distributions $U(a,b)$ MB and $U(c,d)$ MB, respectively. We also assume that each task has required $C_i$ CPU processing cycles defined by $\alpha_i \times D_i^i$ Giga cycles, in which the parameter $\alpha_i$ Giga cycles/MB is the complexity ratio of the task. 
All parameters are given in Table~\ref{tab:Experimental-parameters}.
Specially, the maximum theoretically supported physical-layer data rate of 72 Mbps (the WiFi highest physical-layer data rate of 802.11n smartphones in~\cite{liu2015small,saha2015power}) is used for both the uplink and downlink of each fog node and cloud server $V$.
Here, the energy characteristics of a Nokia N900 with 3G near connection (Table~2~in~\cite{miettinen2010energy}) are used to configure the direct connection between mobile devices and the cloud server $V$. Consequently, the energy consumption for transmitting and receiving a unit of data between mobile devices and the cloud server $V$ are $e_{i(M+1)}^{u} = 0.658$ J/Mb and  $e_{i(M+1)}^{d} = 0.278$ J/Mb, respectively. Besides, the energy consumption for transmitting and receiving a unit of data between mobile device $i$ and fog node $j~(j \leq M)$ are lesser, $e_{ij}^{u} = 0.142$ J/Mb and  $e_{ij}^{d} = 0.142$ J/Mb, respectively (the energy characteristics of a Nokia N900 with WLAN connection in Table~2~in~\cite{miettinen2010energy}).

\begin{table}[htbp]
	\caption{Experimental parameters\label{tab:Experimental-parameters}}
	\centering{}%
	\begin{tabular}{|l|c|}
		\hline 
		\textbf{Parameters} & \textbf{Value}\tabularnewline
		\hline 
		Number of mobile devices $N$ & 10\tabularnewline
		\hline 
		Number of fog nodes $M$ & 4\tabularnewline
		\hline 
		Number of cloud server $V$ & 1\tabularnewline
		\hline		
		CPU rate of mobile devices $f_i^l$ & $0.5$ Giga cycles/s\tabularnewline
		\hline
		Processing energy consumption rate $v_i$ & $\frac{1000}{730}$ J/Giga cycles\tabularnewline
		\hline  
		Input data size $D_i^i$ & $U(a,b)$ MB\tabularnewline
		\hline
		Output data size $D_i^o$ & $U(c,d)$ MB\tabularnewline
		\hline
		Required CPU cycles $C_i$ & $\alpha_i \times D_i^i$ \tabularnewline
		\hline
		Unit transmission energy consumption & $0.142$  J/Mb\tabularnewline
		to fog nodes $e_{ij}^u~(\forall j \leq M)$ & \tabularnewline
		\hline
		Unit receiving energy consumption & $0.142$ J/Mb\tabularnewline
		from fog nodes  $e_{ij}^d (\forall j \leq M)$ & \tabularnewline
		\hline 
		Unit transmission energy consumption & $0.658$  J/Mb\tabularnewline
		to cloud server $V$ $e_{i(M+1)}^u$ & \tabularnewline
		\hline
		Unit receiving energy consumption & $0.278$ J/Mb\tabularnewline
		from cloud server $V$ $e_{i(M+1)}^d$ & \tabularnewline
		\hline 
		Delay requirement $t_i^r$  & $[1,10]$s\tabularnewline
		\hline		
		Processing rate of each fog node $R_j^f$ & $10$ Giga cycles/s\tabularnewline
		\hline
		Uplink data rate of each fog node $R_i^u$ & 72 Mbps\tabularnewline
		\hline 
		Downlink data rate of each fog node $R_i^d$ & 72 Mbps\tabularnewline
		\hline  
		CPU rate of the cloud server $f^c$ & $10$ Giga cycles/s\tabularnewline
		\hline 
		Data rate between FNs and the cloud $w^{c}$ & $5$ Mbps\tabularnewline
		\hline  
	\end{tabular}
\end{table}


Here, we refer the policy in which all tasks are processed locally as ``Without Offloading'' (WOP), and the policy in which all tasks are offloaded to the fog nodes or the cloud server then minimized the average delay of all tasks as the ``All Offloading'' (AOP). 
Due to the simplicities of WOP, AOP, and ROP, these policies are modeled using the GAMS language and solved by the ANTIGONE solver. We develop the proposed algorithms, IBBA and FFBD, as in Algorithm~\ref{IBBA_Algorithm_code}~and~\ref{FFBD_Algorithm_code} using the Optimizer API of the \textbf{MOSEK} solver~\cite{mosek2019documentation}. To evaluate the efficiency of theoretical proposals, each of these methods is implemented with two variants. Specifically, the IBBA variant with the optimal solution selection LFC policy is denoted as IBBA-LFC, whereas IBBA-LCF is the name of the IBBA variant with the LCF policy. In the IBBA-LFC, tasks are branched in the order of mobile device $\rightarrow$ fog nodes $\rightarrow$ cloud server. In the IBBA-LCF, the tasks are branched in the order of mobile device $\rightarrow$ cloud server $\rightarrow$ fog nodes. 
In other words, the IBBA-LFC tries to offload tasks to fog nodes as much as possible before offloading to the cloud server. On the contrary, the IBBA-LCF tries to offload to the cloud server before considering fog nodes. 
Similarly, the FFBD variant using the standard MOSEK solver for subproblems is denoted as FFBD-S, and the one first using the fast solution detection method described in Theorem~\ref{theo:feasible} is denoted as FFBD-F. 
The results obtained by the IBBA-LFC/LCF and FFBD-S/F will be compared with the ROP, WOP, and AOP. To compare the computation time of IBBA-LFC/LCF and FFBD-S/F, the same runtime environment is a normal laptop with Intel Core i5 2.30GHz CPU and 8GB of RAM. Each method runs every experiment $10$ times continuously, then the performance is calculated as the average of these $10$ runs.
Noticeably, the WOP, AOP, and ROP may not satisfy either the delay constraints of tasks or the consumed energy optimization. Thus, all methods will be evaluated the error rates, defined as the proportion of tasks that do not satisfy their delay constraints.

\subsection{Numerical Results}

\subsubsection{Scenario 1 - Vary the Complexity of Tasks} 

In this scenario, we investigate the effect of task complexity on the offloading decisions and energy consumption of mobile devices by varying the complexity of all tasks.

At first, $N$ tasks $I_{i}\left(D_{i}^{i},D_{i}^{o},C_{i},t_{i}^{r}\right)$ are generated as $D_{i}^{i} \sim U(1.0,10.0)$MB,  $D_{i}^{o} \sim U(0.1,1.0)$MB, $C_{i} \sim \alpha_i \times D_i^i$ Giga cycles, and the delay requirement $t_{i}^{r}$ is set to $10s$ for all tasks. Then, the complexity ratio of tasks $\alpha_i$ starts from $U(0.1, 1.0)$ Giga cycles/MB, then increases each task $0.1$ Giga cycles/MB for each experiment. Other parameters are set as in Table~\ref{tab:Experimental-parameters}.

\begin{figure}[!]
	\centering
\includegraphics[height=2in]{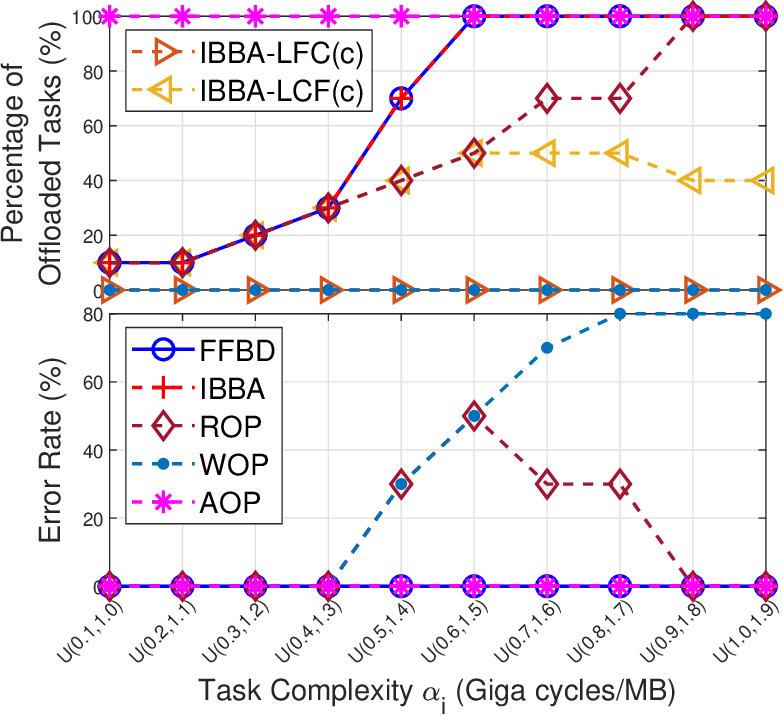}
	\caption{Percentage of offloaded tasks and error rate as the task complexity $\alpha_i$ is increased.}
	\label{fig:scen1_offload_error_rate}
\end{figure}

Fig.~\ref{fig:scen1_offload_error_rate} depicts the percentage of offloaded tasks and error rates, which are the proportion of tasks violating their delay requirements, when the task complexity $\alpha_i$ increases from $U(0.1,1.0)$ to $U(1.0,1.9)$. Generally, while the offloading trends of WOP and AOP are constants, i.e., $0\%$ and $100\%$, respectively, the offloading trends of the FFBD, IBBA, and ROP methods increase dramatically from $10\%$ to $100\%$. This is because there are more tasks benefit from offloading since their complexity $\alpha_i$ is greater than $\alpha_i^*=0.911$ Giga cycles/MB according to Eq.~(\ref{eq:complexity_rate}).
Besides, the percentage of tasks processed at the cloud server (i.e., labeled \textit{IBBA-LFC(c)}) for the IBBA-LFC method is $0\%$ due to the fog processing priority and the sufficient resources at fog nodes. However, only a proportion of offloaded tasks (e.g., $40\%$ in $70\%$ offloaded tasks at $\alpha_i = U(0.5, 1.4)$) are processed at the cloud server (i.e., labelled \textit{IBBA-LCF(c)}) for the IBBA-LCF method when $\alpha_i \geq U(0.5,1.4)$. This is because the cloud processing does not satisfy the delay requirements of all offloaded tasks.
From Fig.~\ref{fig:scen1_offload_error_rate}, the zero error rates indicate the reliability of the FFBD and IBBA methods since their offloading and resource allocation solutions satisfy tasks' delay requirements, $t_i^r=10s$. 
Besides, since $\alpha_i \geq U(0.5,1.4)$, the local execution does not satisfy the delay requirements of all tasks. Thus, the WOP method records an increasing error rate from $30\%$ to $80\%$ for the last six experiments. Noticeably, the error rate of the approximation method, ROP, is proportional to the difference between the offloading rates of ROP and the optimal methods, FFBD and IBBA (e.g., error rate of $30\%$ with $30\%$ offloaded task difference at $\alpha_i = U(0.5,1.4)$). 

\begin{figure*}[!]
	\centering
	\begin{subfigure}[t]{0.4\textwidth}
		\centering
		\includegraphics[height=2in]{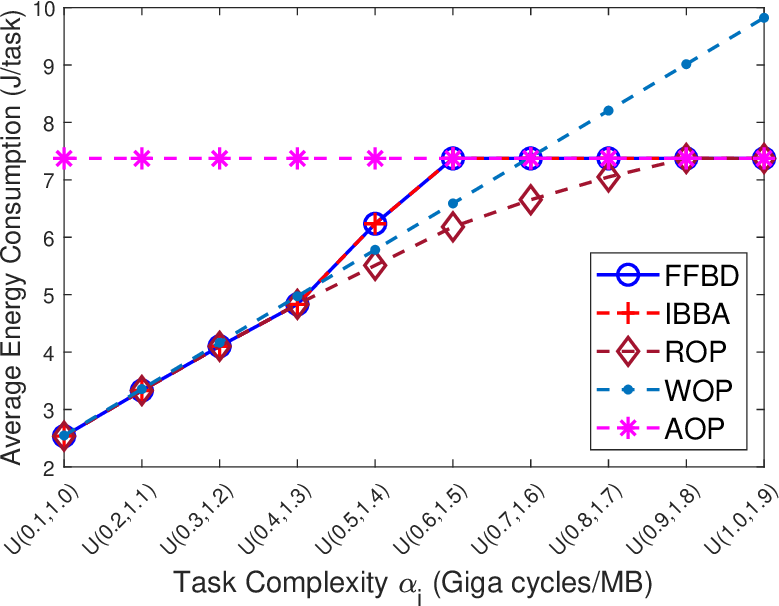}
		\caption{Average consumed energy at mobile devices}
		\label{fig:scen1_energy}
	\end{subfigure}%
	~ 
	\begin{subfigure}[t]{0.4\textwidth}
		\centering
		\includegraphics[height=2in]{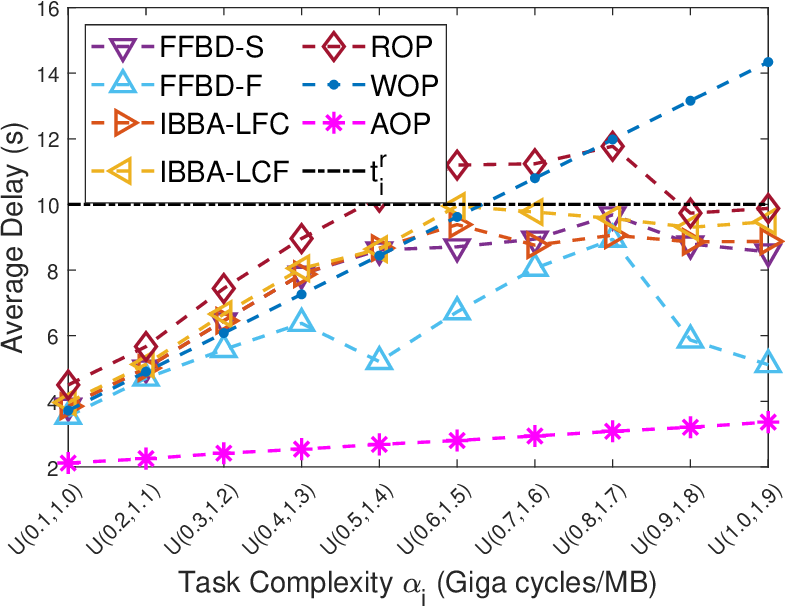}
		\caption{Average task processing delay}
		\label{fig:scen1_delay}
	\end{subfigure}\\
	\caption{Consumed energy and task processing delay as the task complexity $\alpha_i$ is increased.}
	\label{fig:scen1_energy_delay}
\end{figure*}


Fig.~\ref{fig:scen1_energy_delay}(a)~and~\ref{fig:scen1_energy_delay}(b), respectively, show the average consumed energy of mobile devices and delay for the proposed methods when $\alpha_i$ increases from $U(0.1,1.0)$ to $U(1.0,1.9)$ Giga cycles/MB.
Generally, the FFBD and IBBA have the lowest energy consumption in comparing with other methods in all experiments which satisfy the delay requirements. Specifically, due to all offloading without considering energy benefit, the AOP records a constant energy consumption (i.e., $7.37$J/task) which is not lower than that of the FFBD and IBBA methods.
The equality occurs only when $\alpha_i \geq U(0.6,1.5)$ with all tasks being offloaded in the FFBD, IBBA, and AOP methods.
Additionally, Fig.~\ref{fig:scen1_energy_delay}(a) shows that the ROP and WOP can be more energy-efficient at some points in comparing with the FFBD and IBBA, but they must suffer from latency constraint errors as in Fig.~\ref{fig:scen1_offload_error_rate}. For example, when $\alpha_i = U(0.6,1.5)$ the consumed energy of ROP and WOP are $6.2$J/task and $6.6$J/task, respectively, whereas that of the FFBD and IBBA methods is $7.4$J/task. However, the ROP and WOP suffer the equivalent error rates of $50\%$ and $50\%$. Noticeably, all tasks are processed locally in the WOP method, thus the consumed energy increases linearly according to the task complexity ratio.
From Fig.~\ref{fig:scen1_energy_delay}(b), the average delays of the IBBA-LFC/LCF and FFBD-S/F are always lower than the threshold $t_i^r=10$s, matching with their zero error rates.
Additionally, the FFBD-F uses the fast feasible detection method based on Theorem~\ref{theo:feasible}, which allocates the whole communication and computation resources of fog nodes among their assigned tasks.
Consequently, the FFBD-F records the lowest average delay in comparing with the IBBA-LFC/LCF and FFBD-S. Moreover, the fluctuation of delay in the FFBD-F is caused by the different distributions of tasks among fog nodes. 
We also can see the drawback of the ROP since its average delay is bigger than the threshold $t_i^r = 10$s at all experiments with errors.


%


\subsubsection{Scenario 2 - Vary the Task Delay Requirements} 

In this scenario, we study the impact of task delay requirements on the energy consumption of mobile devices and the computation time of the proposed methods.

In this scenario, $N$ tasks $I_{i}\left(D_{i}^{i},D_{i}^{o},C_{i},t_{i}^{r}\right)$ are generated as $D_{i}^{i} \sim U(1.0,10.0)$MB,  $D_{i}^{o} \sim U(0.1,1.0)$MB, $C_{i} \sim \alpha_i \times D_i^i$ Giga cycles, and $t_{i}^{r}$ varying between $(2,10)$s. Besides, we choose a wider complexity rate $\alpha_i=U(0.1, 6.0)$.
After creating the data set, we detect that there are $5$ tasks receiving benefits from offloading $(E_i^l > E_i^f)$ due to $\alpha_i > \alpha_i^*=0.911$ Giga cycles/MB, and all tasks have the local delay between $2$s and $24$s.

\begin{figure*}[!]
	\centering
	\begin{subfigure}[t]{0.4\textwidth}
		\centering
		\includegraphics[height=2in]{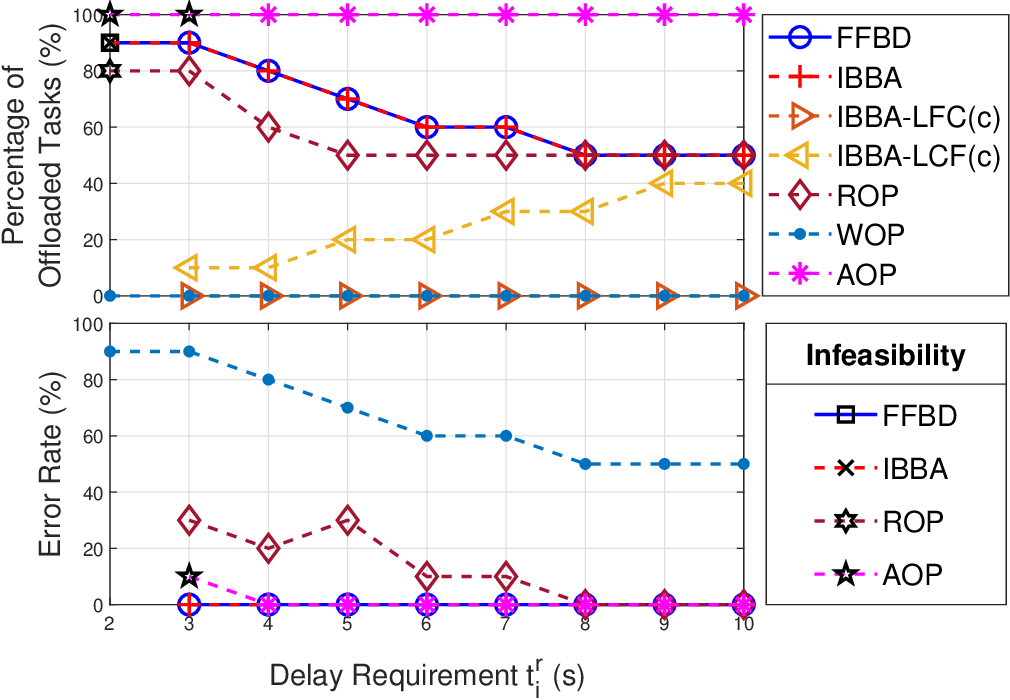}
		\caption{Percentage of offloaded tasks and error rate}
		\label{fig:scen2_offload_error_rate}
	\end{subfigure}%
	~ 
	\begin{subfigure}[t]{0.4\textwidth}
		\centering
		\includegraphics[height=2in]{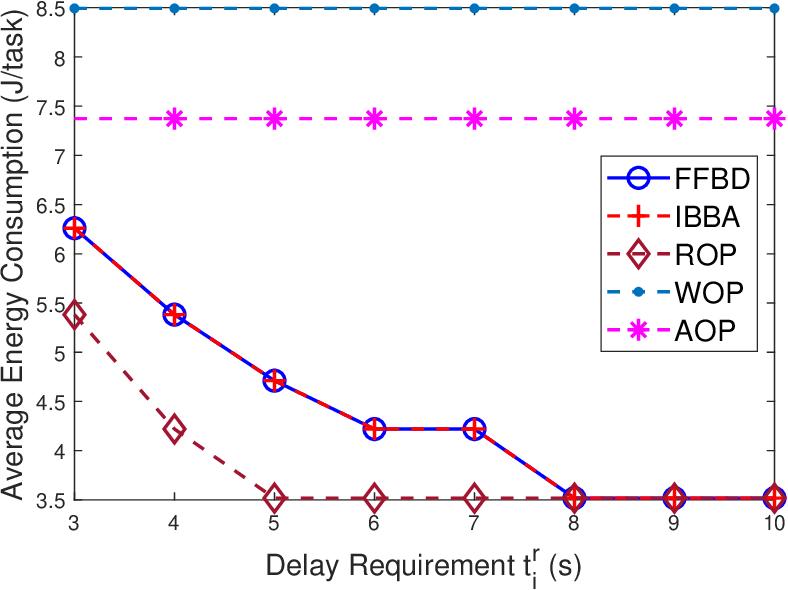}
		\caption{Average consumed energy at mobile devices}
		\label{fig:scen2_energy_required}
	\end{subfigure}
	\caption{Percentage of offloaded tasks, error rate,  and average consumed energy as the delay requirement $t_i^r$ is less strict.}
	\label{fig:scen2_offload_error_rate_energy}
\end{figure*}

Fig.~\ref{fig:scen2_offload_error_rate_energy}(a) shows the offloading trends and error rates when the delay requirement goes up from $2$s to $10$s. Generally, while the trends of WOP and AOP are constants, i.e., $0\%$ and $100\%$, respectively, the offloading trends of the FFBD and IBBA methods decrease from $90\%$ to $50\%$. Specifically, at first some tasks without offloading benefits still have to be offloaded due to their high local processing delay $(T_i^l > t_i^r)$, then when $t_i^r$ is larger, these tasks will be executed locally to reduce the consumed energy if $T_i^l \leq t_i^r$. Noticeably, the fog computing system does not have enough resources to process all tasks satisfying the delay requirement $t_i^r \leq 3$, hence it is infeasible at $t_i^r = 2$s for the FFBD, IBBA and ROP methods, and at $t_i^r=2$s and $3$s for the AOP.
Since $t_i^r \geq 8$s, the FFBD and IBBA return the optimum solution with only $50\%$ offloaded tasks, which get energy benefit from offloading.
Besides, the proportion of tasks processed at the cloud server (i.e., labelled \textit{IBBA-LFC(c)}) for the IBBA-LFC method is $0\%$ due to the fog processing priority and the sufficient resources at fog nodes. 
However, in the IBBA-LCF method, the proportion of offloaded tasks being processed at the cloud server (i.e., labeled \textit{IBBA-LCF(c)}) increases from $10\%$ in $90\%$ to $40\%$ in $50\%$ when the delay requirements $t_i^r$ gradually grows from $3$s to $10$s. This is because the cloud computing can satisfy more tasks with the looser delay thresholds.
From Fig.~\ref{fig:scen2_offload_error_rate_energy}(a), generally, while the error rate of FFBD and IBBA is zero, it is generally decreases for other methods. The WOP records the highest error rate, steadily decreasing from $90\%$ to $50\%$ when the delay requirement is looser. The ROP records non-zero error rates when $T_i^r$ is between $3$s and $7$s.  




The offloading trends completely match with the average energy consumption depicted in Fig.~\ref{fig:scen2_offload_error_rate_energy}(b). Generally, while it is a constant for both WOP with $8.5$J/task and AOP with $7.4$J/task, the consumed energy in the FFBD and IBBA decreases from $6.3$J/task to $3.5$J/task when increasing the delay requirement. Equivalently, both the FFBD and IBBA methods reduce the consumed energy from $15\%$ to $52\%$ and from $26\%$ to $59\%$, respectively, in comparing with AOP and WOP.
Especially, in comparing with the FFBD and IBBA methods, although the ROP method achieves energy benefits at some experiments, it must suffer from latency errors.





\subsubsection{Complexity and Computation Time}

In this subsection, we present some results on the complexity of the four algorithms, IBBA-LFC/LCF and FFBD-S/F, and the necessary time to find an optimal solution in each experiment. Here, the WOP, AOP, and ROP are ignored due to their inadequacy of the goal. 

We also evaluate the efficiency of integrating ROP into the FFBD method. Specifically, the methods, in which the offloading decision and resource allocation solutions of ROP are used as the initial point for FFBD-S/F, are named ROP-FFBD-S and ROP-FFBD-F, respectively. Moreover, the solutions of the master problem and subproblems from previous iterations are also used as the initial points of the current iteration.

The complexity of the four algorithms is calculated by the \textit{number of intermediate problems} (i.e., the intermediate relaxed problems during searching trees in the IBBA-LFC/LCF, master problem $(\mathbf{MP}_0)$ and subproblem $(\mathbf{SP}_1)$ solved by the standard solver in FFBD-S/F). In the FFBD-F, due to the low complexity, the subproblems solved by the fast feasible detection method as in Theorem~\ref{theo:feasible} are ignored.

\begin{figure*}[!]
	\centering
	\begin{subfigure}[t]{0.4\textwidth}
		\centering
\includegraphics[height=2.2in]{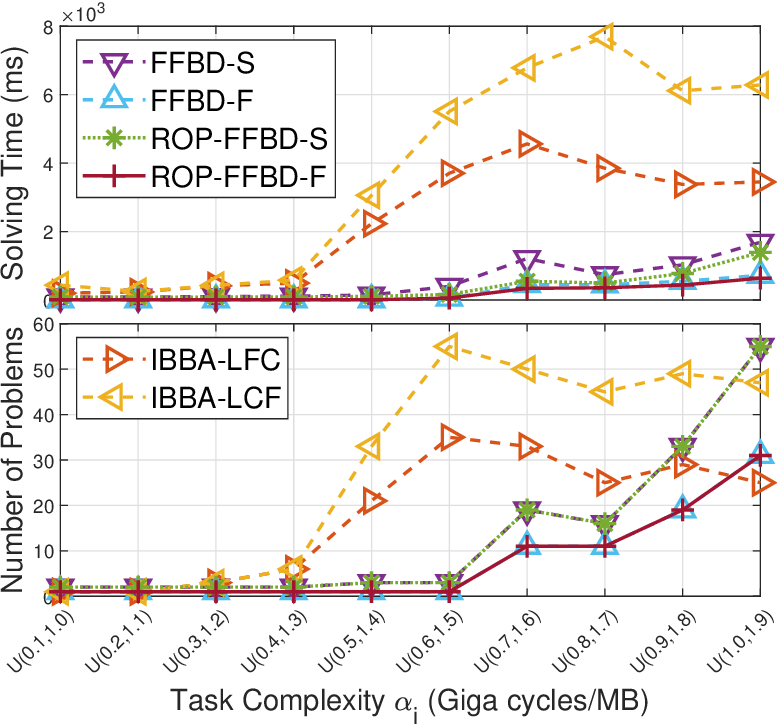}
	\caption{When task complexity $\alpha_i$ is increased}
	\label{fig:scen1_solving_time_num_of_prob}
	\end{subfigure}%
	~ 
	\begin{subfigure}[t]{0.4\textwidth}
		\centering
	\includegraphics[height=2.2in]{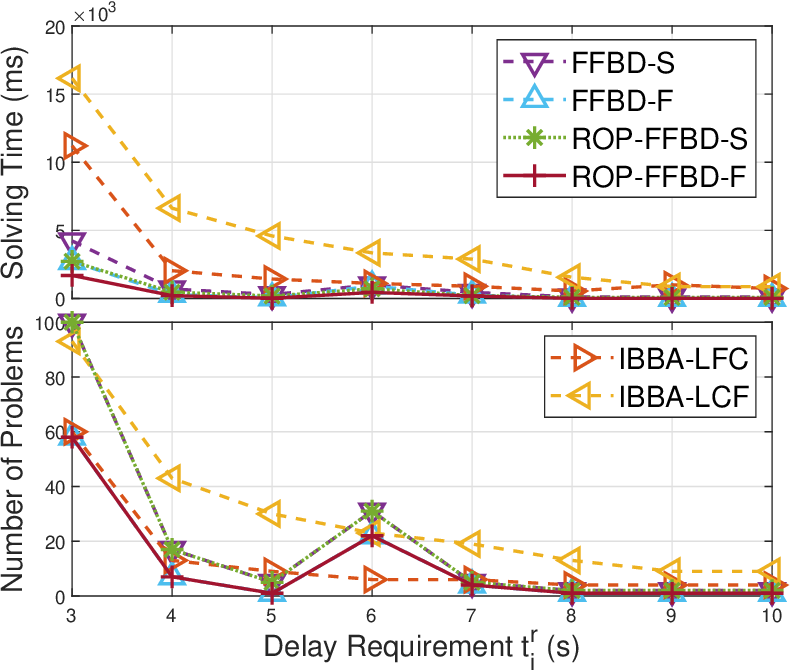}
	\caption{When delay requirement $t_i^r$ is less strict.}
	\label{fig:scen2_solving_time_num_of_prob}
	\end{subfigure}
	\caption{Computation time and number of solved intermediate problems in order to find an optimal solution.}
	\label{fig:solving_time_num_of_prob}
\end{figure*}

Fig.~\ref{fig:solving_time_num_of_prob}(a)~and~\ref{fig:solving_time_num_of_prob}(b) show the computation time and the number of intermediate problems being solved since either the task complexity goes up or the delay requirement is looser.
Generally, the computation time is proportional to the number of intermediate problems. Noticeably, although the FFBD-F/S methods have to solve much more intermediate problems in some experiments, their computation time is still remarkably lower than that of the IBBA-LFC/LCF methods due to the small size of their intermediate problems. This shows the efficiency of the decomposition, initial Benders cuts based on Theorem~\ref{theo:infeasible} and the cutting-plane generation from the results of subproblems in the FFBD. 
Besides, when either tasks have a higher complexity or a lower delay requirement, the FFBD-S/F and IBBA-LFC/LCF methods need to solve more intermediate problems in order to satisfy more resource demands of tasks, thus they require more time to find the optimal solutions. 
Especially, the FFBD-F with the fast solution detection method can reduce, respectively, $60\%$, $90\%$ and $94\%$ of the average computation time in comparing with the FFBD-S, IBBA-LFC and IBBA-LCF methods for Scenario 1, and $40\%$, $78\%$ and $89\%$ for Scenario 2.
Fig.~\ref{fig:solving_time_num_of_prob} also shows that integrating the approximated solution of ROP and intermediate solutions of iterations into FFBD does not reduce the number of problems, but it improves the solving time. Specifically, the solving time of ROP-FFBD-S and ROP-FFBD-F are, respectively, lower than that of FFBD-S and FFBD-F in all experiments.

Table~\ref{tab:computation_time} summarizes the major performance involving the computation time, the standard solver and fast solution detection method usages, and the master problem iterations. For Scenario 1, the ROP-FFBD-F, FFBD-F, ROP-FFBD-S, FFBD-S, IBBA-LFC and IBBA-LCF algorithms, respectively, have the average solving time of $186$ms, $224$ms, $382$ms, $561$ms, $2252$ms, and $3714$ms equivalent to an average of $3.5$, $3.5$, $9.4$, $9.4$, $17.9$, and $29$ times using the standard solver for the subproblems, which are either the intermediate relaxed problems in the IBBA-LFC/LCF or the subproblems of the form $(\mathbf{SP}_1)$ solved by the standard solver in the FFBD-S/F. 
For Scenario 2, the maximum number of master problem iterations is 20 for the FFBD-F/S, and average $8.63$ ($56.1\%$) of $15.38$ subproblems are solved by the fast feasible method for the FFBD-F. 
From Fig.~\ref{fig:solving_time_num_of_prob}(a),~\ref{fig:solving_time_num_of_prob}(b) and Table~\ref{tab:computation_time}, we can conclude that the computation time depends not only on the number and size of intermediate problems but also their specific properties, which correlate with the distance between the intermediate solutions and the optimal one. 

\begin{table*}[!]
	\caption{Complexity and computation times\label{tab:computation_time}}
	\centering{}%
	\begin{tabular}{|l|c|c|c|c|c|c|}
		
		\hline 
		\multicolumn{7}{|l|}{\textbf{Scenario 1: Increasing complexity}}\tabularnewline
		
		\hline 
		& \textbf{ROP-FFBD-F} & \textbf{FFBD-F} & \textbf{ROP-FFBD-S} & \textbf{FFBD-S} & \textbf{IBBA-LFC} &\textbf{IBBA-LCF}\tabularnewline
		
		\hline 
		Min. time & 5ms & 5ms & 86ms & 86ms & 200ms & 248ms \tabularnewline
		
		\hline
		Max. time & 637 & 721ms & 1390 & 1687ms & 4562ms & 7694ms \tabularnewline
		
		\hline
		Average time & 186ms & 224ms & 382ms & 561ms & 2252ms & 3714ms \tabularnewline
		
		\hline
		Average num. of & 3.5 & 3.5 & 9.4 & 9.4 & 17.9 & 29  \tabularnewline
		standard solve & (37.2\%) & (37.2\%) & (100\%) & (100\%) & (100\%) & (100\%)  \tabularnewline
		
		\hline
		Average num. of & 5.90 & 5.90 &NA & NA & NA & NA  \tabularnewline
		fast solve & (62.8\%) & (62.8\%) & (0\%) & (0\%) & (0\%) & (0\%)  \tabularnewline
		
		\hline
		Max. MP & 16 & 16 & 16 & 16 & NA & NA  \tabularnewline
		Iterations & & & &  &  &   \tabularnewline
		
		\hline
		Average MP  & 4.3 & 4.3 & 4.3 & 4.3 & NA & NA  \tabularnewline
		Iterations & & & &  &  &   \tabularnewline
		
		\hline 
		\multicolumn{7}{|l|}{\textbf{Scenario 2: Vary the task delay requirements}}\tabularnewline
		
		\hline 
		& \textbf{ROP-FFBD-F} & \textbf{FFBD-F} & \textbf{ROP-FFBD-S} & \textbf{FFBD-S} & \textbf{IBBA-LFC} &\textbf{IBBA-LCF}\tabularnewline
		
		\hline 
		Min. time & 6ms  & 8ms & 79ms & 111ms & 548ms & 866ms \tabularnewline
		
		\hline
		Max. time & 1692ms & 2715ms & 2726ms & 4224ms & 11206ms & 16160ms \tabularnewline
		
		\hline
		Average time & 321ms & 530ms & 568ms & 878ms & 2371ms & 4615ms \tabularnewline
		
		\hline
		Average num. of & 6.75 & 6.75 & 15.38 & 15.38 & 14.5 & 29.88  \tabularnewline
		standard solve & (43.9\%) & (43.9\%) & (100\%) & (100\%) & (100\%) & (100\%)  \tabularnewline
		
		\hline
		Average num. of & 8.63 & 8.63 & NA & NA & NA & NA  \tabularnewline
		fast solve & (56.1\%) & (56.1\%) & (0\%) & (0\%) & (0\%) & (0\%)  \tabularnewline
		
		\hline
		Max. MP & 20 & 20 & 20 & 20 & NA & NA  \tabularnewline
		Iterations & & & &  &  &   \tabularnewline
		
		\hline
		Average MP & 5.13 & 5.13 & 5.13 & 5.13 & NA & NA  \tabularnewline
		Iterations &  & & &  &  &   \tabularnewline

		\hline 
		
		\hline
	\end{tabular}
\end{table*}


\section{Conclusion} 
\label{sec:conclusion}

We have proposed the joint offloading decision and resource allocation optimization framework for the multi-layer cooperative fog computing network. To find the optimal solution, we have developed three effective methods called IBBA with two variants IBBA-LFC/LCF (based on the Branch and bound), the distributed method, FFBD, with two variants FFBD-S/F (based on the Benders decomposition) and ROP (an approximation policy based on the solution of the relaxed problem). While the IBBA-LFC/LCF and FFBD-S/F can find the optimal solution, the ROP is a suboptimal method with error rates. The FFBD-F implemented the fast feasible detection method is the fasted algorithm in term of the computation time. Whereas, the IBBA-LFC/LCF algorithms with the optimal solution selection strategies can find the optimal solution with most tasks being offloaded to fog nodes and the cloud server, respectively. Numerical results have demonstrated the efficiency in terms of energy consumption reduction of the proposed methods.


\appendices

\section{Proof of Theorem~\ref{theo:convexity}}
\label{sec:theo_convexity}

\begin{proof} From Eqs.~(\ref{eq:local_en}),~(\ref{eq:fog_en}),~(\ref{eq:cloud_en}),~(\ref{eq:direct_cloud_en}),~and~(\ref{eq:total_en}), the objective function, $E=\mathbf{e}^{\top}\mathbf{x}$, is a linear expression of decision variables $\mathbf{x}$ because $\mathbf{e}$ is independent from $\mathbf{x}$  and $\mathbf{r}$. We need to show that all constraints in $(\mathbf{R}_0)$ and $(\mathbf{\widetilde{X}}_0)$ are convex functions. 
	That is, from Eqs.~(\ref{eq:local_delay}),~(\ref{eq:fog_delay}),~(\ref{eq:cloud_delay}),~(\ref{eq:direct_cloud_delay}),~and~(\ref{eq:task_delay_convex}), the delay $T_{i}=\mathbf{h}_i^{\top}\mathbf{y}_{i}$ is the sum of functions, i.e., $x_{i}^{l2}$, $\frac{x_{ij}^{f2}}{r_{ij}^{u}}$, $\frac{x_{ij}^{f2}}{r_{ij}^{d}}$, $\frac{x_{ij}^{f2}}{r_{ij}^{f}}$, $x_{ij}^{c2}$, $\frac{x_{ij}^{c2}}{r_{ij}^{u}}$ and $\frac{x_{ij}^{c2}}{r_{ij}^{d}}$ $\forall j \in \mathbb{M^*}$, with positive coefficients, e.g., $C_i$, $D_i^i$, $D_i^o$, and  $\left(\frac{D_i^i+D_i^o}{w^{c}}+\frac{C_i}{f^c}\right)$. Obviously, functions of the form $x^2$ are convex. We need to prove functions of the form $g(x,r) = \frac{x^2}{r}$ are convex. 
Let $\mathbf{H} = \nabla^2 g(x,r)$ is the Hessian of $g(x,r)$. Then, given an arbitrary vector $\mathbf{v} = (v_1,v_2)$, we have:
\begin{equation}
\mathbf{v}^\top \mathbf{H} \mathbf{v} = \mathbf{v}^\top	\left[	\begin{array}{cc}
\frac{\partial^2 g}{\partial^2 x} & \frac{\partial^2 g}{\partial x \partial r} \\
\frac{\partial^2 g}{\partial r \partial x} & \frac{\partial^2 g}{\partial^2 r}	
\end{array}	\right] \mathbf{v}
=\frac{2}{r} \left(v_1 - v_2 \frac{x}{r} \right)^2.	
\label{eq:Hessian_matrix}
\end{equation}

%

Since the resource allocation variables $r_{ij}^u, r_{ij}^d, r_{ij}^f \geq 0$ (The equality occurs only when $x_{ij}^f, x_{ij}^c = 0$), we have $r \geq 0$. Consequently, we have $\mathbf{v}^\top \mathbf{H} \mathbf{v} \geq 0$. This implies that $\mathbf{H}$ is a positive semidefinite matrix, and thus $g(x,r)$ is a convex function w.r.t. $(x,r)$~\cite{Boyd2004Convex}. Thus, $T_{i}$ is a convex function since it is the nonnegative weighted sum of convex functions. 
In other words, the constraint $(\mathcal{C}_1)$ in $(\mathbf{R}_0)$ is a convex function w.r.t. $\mathbf{x}$ and $\mathbf{r}$. Besides, the constraints $(\mathcal{C}_2)$, $(\mathcal{C}_3)$, $(\mathcal{C}_4)$, and $(\mathcal{C}_5)$ in $(\mathbf{R}_0)$ and $(\mathbf{\widetilde{X}}_0)$ are linear functions. 

Since the objective function in Eq.~(\ref{eq:total_en}) is a linear function, and all constraints in $(\mathbf{R}_0)$ and $(\mathbf{\widetilde{X}}_0)$ are convex functions, the relaxed problem $(\mathbf{\widetilde{P}}_0)$ is a convex optimization problem~\cite{Boyd2004Convex}.
\end{proof}


\section{Proof of Theorem~\ref{theo:ffbdstop}}
\label{sec:theo_ffbdstop}


\begin{proof}
	We assume the cutting-plane sets of $(\mathbf{MP}_0)$ at iterations $(k)$ and $(k+1)$ are $cuts^{(k)}$ and $cuts^{(k+1)}$, respectively. 
	At iteration $k$, assume $(\mathbf{MP}_0)$ is feasible, and there is at least one infeasible subproblem~$(\mathbf{SP}_1)$.
	Consequently, we have $cuts^{(k)} \subset cuts^{(k+1)}$.
	This leads to $\underset{\mathbf{x} \in \mathbf{X}_0}{\min} \{\mathbf{c}^\top \mathbf{x} \}$ s.t. $cuts^{(k)}$ $\leq$ $\underset{\mathbf{x} \in \mathbf{X}_0}{\min} \{\mathbf{c}^\top \mathbf{x} \}$ s.t. $cuts^{(k+1)}$. In other words, $\underset{\mathbf{x} \in \mathbf{X}_0}{\min} \{\mathbf{c}^\top \mathbf{x}\}$ s.t. $cuts^{(k)}$ is a function that does not decrease with iteration $k$.
	Therefore, the first found feasible solution $(\mathbf{x},\mathbf{r})$ of $(\mathbf{MP}_0)$ and $(\mathbf{SP}_0)$ is the optimal solution of~$(\mathbf{P}_0)$.
	
	In the case that $(\mathbf{MP}_0)$ is infeasible at iteration $k$, it means that $(\mathbf{MP}_0)$ will be infeasible at all later iterations due to $cuts^{(k)} \subset cuts^{(k+v)}, \forall v \geq 1$. In other words, the original problem $(\mathbf{P}_0)$ is infeasible.
\end{proof}

%
%


\section{Proof of Lemma~\ref{lem:inequality}}
\label{sec:lem_inequality}

\begin{proof}
	if $\frac{p_1}{q_1} \geq \frac{p_2}{q_2}$ then $\max\{\frac{p_1}{q_1}, \frac{p_2}{q_2}\} = \frac{p_1}{q_1} \geq \frac{p_1+p_2}{q_1+q_2}$. Otherwise, if $\frac{p_1}{q_1} < \frac{p_2}{q_2}$ then $\max\{\frac{p_1}{q_1}, \frac{p_2}{q_2}\} = \frac{p_2}{q_2} > \frac{p_1+p_2}{q_1+q_2}$. In other words, $\max\{\frac{p_1}{q_1}, \frac{p_2}{q_2}\} \geq \frac{p_1+p_2}{q_1+q_2}$. 
	Similarly, $\max\{\frac{p_1+p_2}{q_1+q_2}, \frac{p_3}{q_3}\} \geq \frac{p_1+p_2+p_3}{q_1+q_2+q_3}$. Therefore, $\max\{\frac{p_1}{q_1}, \frac{p_2}{q_2}, \frac{p_3}{q_3}\} \geq \max\{\frac{p_1+p_2}{q_1+q_2}, \frac{p_3}{q_3}\} \geq \frac{p_1+p_2+p_3}{q_1+q_2+q_3}$. 
	Repeatedly, we have $\underset{i \in N}{\max} \{\frac{p_i}{q_i}\} \geq \frac{P}{Q}$.
\end{proof}

\section{Proof of Theorem~\ref{theo:feasible}}
\label{sec:theo_feasible}
\begin{proof}
	Let's find a feasible solution of $(\mathbf{SP}_1)$. Task $I_i$ will be allocated with resources $r_{ij}^u$, $r_{ij}^d$ and $r_{ij}^f$ as
	$r_{ij}^u = \frac{D_i^{i'}}{\beta_{bal}^u}$, $r_{ij}^d = \frac{D_i^{o'}}{\beta_{bal}^d}$, and $r_{ij}^f =\frac{C_i^{'}}{\beta_{bal}^f}$.
	We have $\beta_i =
	\frac{D_i^{i'}}{r_{ij}^u}+ \frac{D_i^{o'}}{r_{ij}^d}+
	\frac{C_i^{'}}{r_{ij}^f} = 
	\left(\beta_{bal}^u + \beta_{bal}^d + \beta_{bal}^f\right)$.
	Here, $r_{ij}^{f} = 0$ and $\frac{C_i^{'}}{r_{ij}^{f}} = 0, \forall i \in \mathbb{N}_j^s.$
	Thus, $\beta_i = \beta_{bal} \leq 1, \forall i \in \mathbb{N}_j^{t+s}$.
	
	Besides, $\sum_{i \in \mathbb{N}_j^{t+s}}r_{ij}^u = R_j^u$, $\sum_{i \in \mathbb{N}_j^{t+s}}r_{ij}^d = R_j^d$ and $\sum_{i \in \mathbb{N}_j^{t+s}}r_{ij}^f = R_j^f$ satisfying resource limit conditions. In conclusion, the problem $(\mathbf{SP}_1)$ is feasible.
\end{proof}

\section{Proof of Theorem~\ref{theo:infeasible}}
\label{sec:theo_infeasible}
\begin{proof}
	Applying Lemma~\ref{lem:inequality} into $\{D_i^{i'}\}_{i \in \mathbb{N}_j^{t+s}}$ and $\{r_{ij}^{u}\}_{i \in \mathbb{N}_j^{t+s}}$, we have $\underset{i \in \mathbb{N}_j^{t+s}}{\max} \{\frac{D_i^{i'}}{r_{ij}^{u}}\} \geq \frac{\sum_{i \in \mathbb{N}_j^{t+s}}{D_i^{i'}}}{\sum_{i \in \mathbb{N}_j^{t+s}}r_{ij}^{u}}$. 	
	According to resource allocation conditions, $\sum_{i \in \mathbb{N}_j^{t+s}}r_{ij}^{u} \leq R_j^u$, we have $\underset{i \in \mathbb{N}_j^{t+s}}{\max} \{\frac{D_i^{i'}}{r_{ij}^{u}}\} \geq \frac{\sum_{i \in \mathbb{N}_j^{t+s}}{D_i^{i'}}}{R_j^u}$. Therefore, $\underset{i \in \mathbb{N}_j^{t+s}}{\max} \{\frac{D_i^{i'}}{r_{ij}^{u}}\} > 1$. Without loss of generality, we assume $\exists i* \in \mathbb{N}_j^{t+s}, \frac{D_{i*}^{i'}}{r_{i*j}^{u}} = \underset{i \in \mathbb{N}_j^{t+s}}{\max} \{\frac{D_i^{i'}}{r_{ij}^{u}}\} > 1$. Consequently, $\beta_{i*} =
	\left(
	\frac{D_{i*}^{i'}}{r_{i*j}^{u}}+
	\frac{D_{i*}^{o'}}{r_{i*j}^{d}}+
	\frac{C_{i*}^{'}}{r_{i*j}^{f}}
	\right) > \frac{D_{i*}^{i'}}{r_{i*j}^{u}} > 1$.
	It contradicts the delay requirement of Task $I_{i*}$, $\beta_{i*} \leq 1$ as in  Eq.~(\ref{eq:delay_constraints_new1}). In conclusion, the problem $(\mathbf{SP}_1)$ is infeasible.
	
	The cases $\frac{\sum_{i \in \mathbb{N}_j^{t+s}}
		D_i^{o'}}{R_j^d} > 1$ and  $\frac{\sum_{i \in \mathbb{N}_j^{t+s}}
		C_i^{'}}{R_j^f} > 1$ are proved in the similar way.
\end{proof}


{\renewcommand{\baselinestretch}{1}
\bibliographystyle{IEEEtran}

\bibliography{IEEE_TCOM_Revision_August2020}}

\end{document}